\documentclass{JHEP3}
\usepackage[english]{babel}
\usepackage{amssymb}
\usepackage{amsfonts}
\usepackage{amsmath}
\usepackage{graphicx}
\usepackage{epsfig}
\usepackage{cite}
\usepackage{psfrag}
\usepackage{subfigure}
\newcommand{\be}{\begin{equation}} \newcommand{\ee}{\end{equation}}
\newcommand{\bea}{\begin{eqnarray}} \newcommand{\eea}{\end{eqnarray}}
\newcommand{\beann}{\begin{eqnarray*}}  \newcommand{\eeann}{\end{eqnarray*}}
\newcommand{\bfig}{\begin{figure}} \newcommand{\efig}{\end{figure}}
\newcommand{\ba}{\begin{array}} \newcommand{\ea}{\end{array}}
\newcommand{\bcen}{\begin{center}} \newcommand{\ecen}{\end{center}}
\newcommand{\btab}{\begin{tabular}} \newcommand{\etab}{\end{tabular}}

\newtheorem{Proposition}{Proposition}[section]

\newtheorem{Theorem}{Theorem}[section]
\newtheorem{Lemma}{Lemma}[section]
\newtheorem{Corrolary}{Corrolary}[section]

\newcommand{\bp}{\begin{Proposition}}	\newcommand{\ep}{\end{Proposition}}
\newcommand{\bt}{\begin{Theorem}}	\newcommand{\et}{\end{Theorem}}
\newcommand{\bl}{\begin{Lemma}}		\newcommand{\el}{\end{Lemma}}
\newcommand{\bc}{\begin{Corrolary}}	\newcommand{\ec}{\end{Corrolary}}

\title{Holographic Evolution of Entanglement Entropy}
\author{Javier Abajo-Arrastia, Jo\~ao Apar{\'\i}cio, Esperanza L\'opez\\
   Instituto de F\'{\i}sica Te\'orica CSIC/UAM\\
   Facultad de Ciencias, modulo C-8\\
   Universidad Aut\'onoma de Madrid\\
  28049 Madrid, Spain\\
  E-mail: \email{javier.abajo, esperanza.lopez@uam.es, jpmn.aparicio@gmail.com}\\
}
\abstract{We study the evolution of entanglement entropy in a 2-dimensional equilibration process that has a holographic description in terms of a Vaidya geometry. It models a unitary evolution in which the field theory starts in a pure state, its vacuum, and undergoes a perturbation that brings it far from equilibrium. The entanglement entropy in this set up provides a measurement of the quantum entanglement in the system. Using holographic techniques we recover the same result obtained before from the study of processes triggered by a sudden change in a parameter of the hamiltonian, known as quantum quenches. Namely, entanglement in 2-dimensional conformal field theories propagates with velocity $v^2\!=\!1$  \cite{Calabrese:2005in}. Both in quantum quenches and in the Vaidya model equilibration is only achieved at the local level. Remarkably, the holographic derivation of this last fact requires information from behind the apparent horizon generated in the process of gravitational collapse described by the Vaidya geometry. In the early stages of the evolution the apparent horizon seems however to play no relevant role with regard to the entanglement entropy. We speculate on the possibility of deriving a thermalization time for occupation numbers from our analysis.

}

\preprint{IFT-UAM/CSIC-10-34} 
\keywords{Holography, out of equilibrium field theory}

\begin{document}
%
\section{\label{sec:intro}Introduction}

The AdS/CFT correspondence has proven to be a unique tool for the study of strongly coupled quantum field theories. Since its first proposal a fruitful subject of application has been thermodynamics, where the holographic dictionary relates the plasma phase of strongly coupled field theories at thermal equilibrium to black hole geometries \cite{Witten:1998zw}. In the last years the correspondence has been generalized to describe small deviations out of equilibrium in the context of linear response theory, and especially its hydrodynamic long wavelength limit \cite{Son:2007vk}. It turns out that holographic plasmas are close to perfect fluids, with very low specific viscosity \cite{Kovtun:2004de}. This fact describes well the data from heavy ion experiments and has represented a major success for the holographic techniques, which were extended in \cite{Baier:2007ix,Bhattacharyya:2008jc} to the study of non-linear fluid dynamics.

Continuing this line of work, it is very interesting to apply holography to the analysis of strongly coupled field theories far from equilibrium. On general grounds, a system brought into a highly excited state is expected to evolve towards a stationary state that at the macroscopical level can be described as thermal equilibrium. The holographic counterpart of this process should correspond to a process of gravitational collapse ending in the formation of a black hole \cite{Banks:1998dd,Danielsson:1999zt,Danielsson:1999fa}. Recently, the far from equilibrium evolution of an anisotropic plasma \cite{Chesler:2008hg} and a boost invariant plasma \cite{Beuf:2009cx,Chesler:2009cy} have been addressed using holography. In \cite{Bhattacharyya:2009uu} the collapse of a shell build out of an AdS massless scalar was analyzed in the limit of weak fields, discussing its implications for field theory thermalization at strong coupling. Time-dependent backgrounds associated with a localized sector of the dual field theory have been constructed in the probe approximation in \cite{Das:2010yw}. Articles \cite{Chesler:2008hg,Beuf:2009cx,Chesler:2009cy,Bhattacharyya:2009uu} mostly considered field theory observables that can be extracted from an expansion of the dual gravity background close to the boundary, in particular the energy-momentum tensor. We want instead to follow the thermalization process using a quantity for which this information is not enough. 

Entanglement is a fundamental notion in quantum physics. A measure of quantum entangled in extended systems is provided by entanglement entropy. The reduced density matrix of a subsystem $A$ is obtained by tracing over the degrees of freedom of its complementary, $\rho_A\!=\!{\rm Tr}_{\bar A} \, \rho$, where $\rho$ is the density matrix of the entire system. The entanglement entropy of region $A$ is defined as the von Neumann entropy of this reduced density matrix
\be
S_A=-{\rm Tr}_A \, ( \rho_A \,{\rm ln} \, \rho_A) \, .
\label{ee}
\ee
When the system is in a pure state, $\rho=|\Psi\rangle \langle \Psi |$, it gives a measure of the quantum entanglement between the degrees of freedom inside and outside $A$. In this situation the von Neumann entropy of the complete system is zero and the following property is fulfilled
\be
S_A=S_{\bar A} \, .
\label{aabar}
\ee
When the system is instead in a mixed state, the entanglement entropy receives also statistical contributions. 

The study of the entanglement entropy is a subject of active research. Many results are available for 2-dimensional field theories and quantum spin chains, see \cite{Calabrese:2009qy} for a review and references. Dynamical situations have been also considered, in particular that triggered by a sudden change in a parameter of the hamiltonian \cite{Calabrese:2005in}. This action, known as quantum quench, can be realized experimentally in systems of cold atoms, together with a subsequent time development under negligible dissipation. This has provided a strong motivation for the study of quantum quenches. In this set up, the system is prepared in a pure state before the quench, $|\Psi_0\rangle$, and due to the unitary character of the subsequent evolution it remains in a pure state, $|\Psi(t)\rangle$. As mentioned above, under this condition the entanglement entropy will describe the evolution of quantum entanglement in the system. General results were obtained in \cite{Calabrese:2005in} for 2-dimensional conformal field theories. It was shown that entanglement propagates from small to large scales with velocity $v^2\!=\!1$. This has the important consequence that equilibration can only be achieved at the local level.

We will address the evolution of entanglement entropy using holographic techniques. The AdS/CFT correspondence has proven to be an efficient method for the calculation of the entanglement entropy in those field theories that admit a gravity dual, see \cite{Nishioka:2009un} for a review. The initial proposal for the holographic derivation of entanglement entropy in static backgrounds \cite{Ryu:2006bv,Ryu:2006ef}, based on the calculation of minimal surfaces, was generalized to time dependent situations in \cite{Hubeny:2007xt}. We will focus on the duality between 2-dimensional CFT's and gravity on asymptotically AdS$_3$ geometries.  
In this case the holographic proposal relates the entanglement entropy with the length  of spacelike geodesics that anchor on the AdS boundary. We will use a Vaidya geometry, which describes the formation of a black hole out of a collapsing shell of null dust, as dual model for a field theory equilibration process. The Vaidya metrics are analytic solutions of Einstein equations that will allow us to carry out a detailed analysis of the relevant geodesics. 

This holographic model realizes a similar set up as the quantum quench. The system starts in a pure state, in this case the vacuum of the dual CFT, and undergoes a perturbation, modeled by the collapsing shell, which brings it into a highly excited state. We will argue that \eqref{aabar} is verified along the complete evolution, implying that the evolution is unitary and the system always remains in a pure state. The Vaidya model and the quantum quench only differ in the type of perturbation that triggers the time development, which determines the initial entanglement pattern. While for the quenches considered in \cite{Calabrese:2005in} entanglement is initially localized on very small scales, in the Vaidya model there are long range correlations in the early stages of the evolution. However we will recover the same conclusions about the evolution of entanglement in 2-dimensional CFT's. This offers a new and strong test of the AdS/CFT correspondence, and at the same time provides a new example where a detailed analysis of the time-development can be performed. A holographic model for a quantum quench that only affects some localized degrees of freedom has been proposed in \cite{Das:2010yw}.

An interesting question for the AdS/CFT correspondence is the possible role of generalized notions of horizon\cite{Ashtekar:2004cn,Booth:2005qc}, like the apparent horizon, in the holographic description of field theory dynamics far from equilibrium. The Vaidya model will allow us to explore 
the relevance of the apparent horizon for the description of the dual field theory entanglement properties. Contrary to the case of the event horizon in static geometries, we will see that the apparent horizon does not act as a wall for the geodesics relevant to the derivation of the entanglement entropy. As happened for quantum quenches, our results show that relaxation is only achieved at the local level. Interestingly geodesics able to detect this effect reach behind both the event and the apparent horizons. Moreover, in the first stages of the evolution the geodesics do not appear to feel the location of the apparent horizon. Hence the apparent horizon seems to play a weak role for the description of quantum entanglement in the far from equilibrium field theory evolution.

The organization of the paper is as follows. Section 2 contains a brief summary about entanglement entropy in 2-dimensional CFT's and its evolution after a quantum quench, together with the holographic proposal for its evaluation. In section 3 we introduce the Vaidya geometry. 
In section 4 we present the numerical study of spacelike geodesics that anchor on the boundary of a 3-dimensional AdS-Vaidya geometry, with emphasis on the role played by the apparent horizon. The implications for the evolution of entanglement are analyzed in section 5, comparing with the known results for quantum quenches. We end in section 6 with a further discussion, where we comment on the possibility of deriving a thermalization time for occupation numbers from our results.

\section{\label{entent}Entanglement entropy}

A remarkable property of entanglement entropy is its universal behavior in 2-dimensional conformal field theories \cite{Holzhey:1994we,Vidal:2002rm,Latorre:2003kg,Korepin:2003,Calabrese:2004eu}. In the ground state of a CFT living on a line, the entanglement entropy of an interval scales logarithmically with its size
\be
S(l)={c \over 3}\, {\rm log} \,{l \over \epsilon} \, ,
\label{eevac}
\ee
where $\epsilon$ is a UV cutoff. The only information relevant to $S(l)$ is the CFT central charge, $c$, a quantity that provides a measurement of the number of degrees of freedom in the theory. In \eqref{eevac}  we have omitted an $l$-independent non-universal contribution. Up to the same constant, the entanglement entropy of an interval at thermal equilibrium is given by the following expression \cite{Calabrese:2004eu}
\be
S_\beta(l)= {c \over 3}\,{\rm log} \left( {\beta \over \pi \epsilon} \,{\rm sinh} \,{\pi l \over \beta} \right) \, ,
\label{eethermal}
\ee
where $\beta\!=\!1/T$ is the inverse of the temperature.
The ground state result is recovered for intervals much smaller than $\beta$. For larger intervals the cutoff independent part of the entanglement entropy behaves as an extensive quantity and leads to the same result as the thermal entropy.

The evolution of entanglement entropy in 2-dimensional systems after a quantum quench was analyzed in \cite{Calabrese:2005in}. In this set up the system is prepared in its ground state, and at $t\!=\!0$ a parameter of the hamiltonian is suddenly changed. The old ground state provides a highly excited starting point for the evolution with respect to the new hamiltonian. 
Explicit results can be obtained for 2-dimensional field theories when the ground state previous to the quench has a mass gap while the dynamics after the quench is conformal. For large times and intervals, $t,l\!>\!>\!\tau_0$ with $\tau_0$ of the order of the inverse mass gap in the initial state, the entanglement entropy of an interval is given by \cite{Calabrese:2005in,Calabrese:2007rg,Calabrese:2009qy}
\be
S(l,t) \, \simeq \, {c \over 3} \,{\rm ln} {2\tau_0 \over \pi}\, + \, \left\{ 
\begin{array}{cc}
\displaystyle \frac{\pi c t} {6 \tau_0} &\hspace{1cm} t<l/2 \, , \\[6mm]
\displaystyle {\pi c l \over 12 \tau_0} &\hspace{1cm} t>l/2 \, .
\end{array}
\right.
\label{eeevol}
\ee
The sharp cusp in this expression at $t\!=\!l/2$ should be understood as smoothed out over a region of order $\tau_0$. After the quench $S(l,t)$ grows linearly with time until it saturates at $t\!\simeq\!l/2$. Comparison with \eqref{eethermal} shows that for late times the entanglement entropy behaves as that of a thermal state, with $4\tau_0$ playing the role of the inverse temperature. An analogous conclusion is derived from the evaluation of correlation functions \cite{Calabrese:2006rx,Calabrese:2007rg}. This leads to interpret $1/4 \tau_0$ as an effective temperature for the long wavelength modes \cite{Calabrese:2007rg}.

\begin{figure}[thp]
\begin{center}
\includegraphics[scale=0.6]{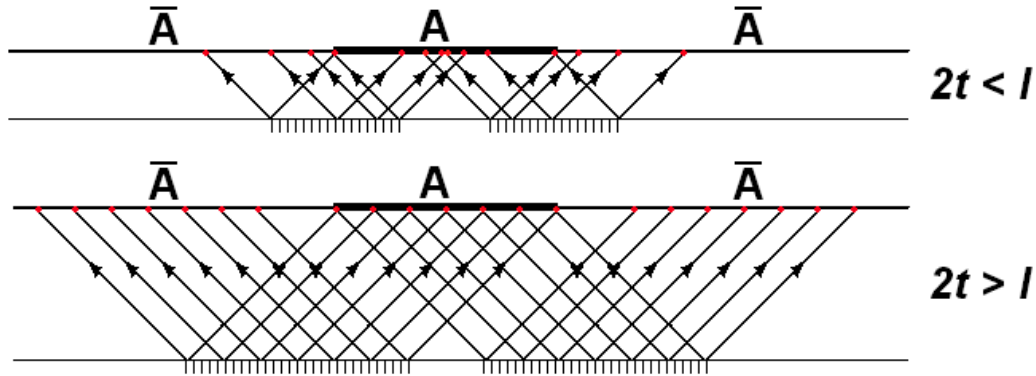}
\end{center}
\caption{\label{fig:prop} Derivation of the entanglement entropy of an interval according to the model proposed in \cite{Calabrese:2005in}. Only excitations originating from the shaded region at $t\!=\!0$ contribute to the entanglement entropy of region $A$ at time $t$.}
\end{figure}

The physical interpretation of \eqref{eeevol} is as follows.
The non-zero energy density created at $t\!=\!0$ by the quench translates into the production of a translationally invariant sea of excitations. Entangled excitations will tend to spread over ever larger regions mainly as a result of propagation. Since the ground state before the quench has a mass gap and hence a finite correlation length, only excitations produced approximately at the same point will be entangled. A simple picture where entangled excitations are assumed to move classically and without scattering, was further proposed in \cite{Calabrese:2005in}. 
For the critical quench described above, their velocity is taken to be $v^2\!=\!1$ equal for all momentum modes. At time $t$ left and right-moving entangled excitations will have separated a distance $2t$. In order to contribute to the entanglement entropy, only the left-moving or the right-moving components must lie inside the chosen interval, see Fig.\ref{fig:prop}. When the separation reached by entangled excitations is smaller than the size of the interval, the number of them that contributes to $S(l,t)$ grows with time proportional to $4t$. When instead $2t\!>\!l$ no left-right moving entangled excitations are contained inside the interval and $S(l,t)$ saturates at a value proportional to $2l$. Hence, although causality prevents the system to equilibrate at the global level, any finite region will locally relax to a stationary state for which its complementary region acts as a thermal bath.

The evolution of the entanglement entropy after a quantum quench has also been studied in spin chain models \cite{Calabrese:2005in,DeChiara:2005wb,Eisert:2006zz,Bravyi:2006zz,Schuch:2008zza}. As in the previous case, it was found in several examples that the entanglement entropy of a block of spins grows linearly with time until it saturates at a value set by the number of spins in the block.
However the way of approaching the saturation value shows that, while there is a maximal velocity for signal propagation, in general not all modes responsible for the propagation of entanglement reach it. The evolution \eqref{eeevol}, modeled in Fig.\ref{fig:prop}, implies instead that in a 2-dimensional CFT the entanglement pattern of the excited state created by the quench is propagated by all modes with maximal velocity, $v^2\!=\!1$.

The evaluation of the entanglement entropy is a very complicated problem in general situations, since it involves performing traces of the density matrix. Therefore having alternative tools to address this problem is very valuable. This has been provided in the last years by the AdS/CFT correspondence for those field theories that admit a gravity dual \cite{Ryu:2006bv,Ryu:2006ef}. 
The proposal referred to static situation, where there exists a timelike Killing vector in terms of which to define a foliation of the dual geometry by slices of constant time. Restricting to such a slice, the entanglement entropy of a region $A$ is calculated holographically as the area of the minimal surface $\gamma$  that anchors at $\partial A$, the boundary of $A$, at the boundary of the dual gravity background 
\be
S_A={Area(\gamma) \over 4 G_{N}} \, ,
\label{holents}
\ee
with $G_{N}$ the bulk Newton constant. This proposal has been generalized to time dependent situations, where there is no unique way to extend a constant time slice of the boundary into the bulk geometry. It has been argued however that \eqref{holents} holds, interpreting $\gamma$ as an extremal surface in the unrestricted gravity background subject to the same boundary conditions \cite{Hubeny:2007xt}. 

In the absence of other scales than the temperature, the dual gravity description of a field theory in thermal equilibrium is given by a black hole geometry \cite{Witten:1998zw}. We want to study the dynamics of field theories far from equilibrium using holographic techniques. On general grounds the field theory will tend, as time evolves, to a state that at the macroscopical level can be described as thermal equilibrium. The candidate gravity duals should therefore represent a process of gravitational collapse ending in the formation of a black hole \cite{Banks:1998dd,Danielsson:1999zt,Danielsson:1999fa}. There is a family of exact solutions to the Einstein equations which describes the collapse of  null dust to form a black hole, known as Vaidya metrics \cite{Stephani:2003tm}. They were derived for asymptotically flat spaces, but can be generalized to spaces with a cosmological constant. In this paper we will focus on an equilibration process that can be described holographically in terms of a Vaidya geometry. 

More general collapse processes, triggered by an infalling shell built up from the non-normalizable mode of an AdS massless scalar were considered in \cite{Bhattacharyya:2009uu}. According to the holographic dictionary, they describe the perturbation of the dual conformal field theory out of its vacuum state by switching on a time-dependent coupling to a marginal operator. 
The solution of Einstein equations in that case could not be obtained in a close form, being approached instead in terms of a perturbative expansion in the limit of weak perturbations. 
Since null dust models incoherent massless radiation, the Vaidya geometry can be expected to represent a limiting case of the previous set up. Constructing a holographic dual for a quantum quench would require a different dual gravity background, which includes an IR cutoff able to represent a mass gap in the field theory previous to the quench. Therefore the Vaidya geometry offers us the possibility to study the evolution of entanglement entropy in a different dynamical process to that generated by a quantum quench. 

\section{\label{vmetric}Vaidya geometry}

The Vaidya metrics in an asymptotically AdS spacetime of dimension $d\!+\!1$, in Poincar\'e coordinates are given by 
\be
ds^2=-\left( r^2-{m(v)\over r^{d-2}} \right) dv^2+2 dr dv+r^2 \sum_{i=1}^{d-1} dx_i^2 \, .
\label{vaidya}
\ee
They are solutions of Einstein's equations in the presence of the following energy-momentum tensor
\be
T_{vv}=(d-1){1\over 2 r^{d-1}} {d m \over dv} \, ,
\ee
with all other components being equal to zero. The energy-momentum tensor characteristic of null dust satisfies $T_{\mu \nu}\!\propto\! k_\mu k_\nu$ with $k^2\!=\!0$. The previous expression is of this form with $k_\mu\!=\!\delta_{\mu v}$, which generates infalling radial null geodesics. The metric \eqref{vaidya} describes the formation of a black hole out of collapsing null dust, or more in general, the absorption of null dust by a black hole.
The Vaidya metrics were first derived as a model for the exterior geometry of a radiating star \cite{vaidya1,vaidya2}, being formulated in that case in terms of an outgoing null coordinate. 

In the study of gravitational collapse, a central question is how to characterize the formation of a black hole at the level of the local time evolution. The event horizon, being a global property of the spacetime, is not appropriate for this purpose. Several generalized notions of horizon have been proposed based on identifying black holes with regions containing trapped surfaces \cite{Ashtekar:2004cn,Booth:2005qc}. A trapped surface is a closed spacelike surface of co-dimension two for which the expansions along the two future directed normal null directions are negative.
These generalized notions of horizon have in general the drawback of requiring a choice of spacelike surfaces and therefore not being covariantly defined. For the Vaidya metrics a natural choice are surfaces that respect the translational invariance along the $x_i$ coordinates.  
With this choice most proposals coincide. One of them is the apparent horizon, defined as the boundary of trapped surfaces associated to a given foliation \cite{Hawking:1973uf}. Under some smoothness conditions, the apparent horizon on each leaf of the foliation coincides with the co-dimension two surface for which the expansion of outgoing future directed normal null geodesics is zero, while that of ingoing ones contracts. A foliation preserving invariance under translations on the $x_i$ coordinates selects the surfaces $r\!=\!const$, $v\!=\!const$.\footnote{These surfaces could be rendered closed by compactifying the $x_i$ coordinates.}. Their associated normal null vectors are
\be
N_{in}=-\partial_r \, , \hspace{1cm} N_{out}=\partial_v+{1 \over 2}\left(r^2-{m(v)\over r^{d-2}} \right) \partial_r
\ee 
whose normalization has been chosen such that $N_{in}\cdot N_{out}=-1$. The expansion along 
both null directions is given by $\theta = {\tilde g}^{\mu \nu} \nabla_\mu N_\nu$, where ${\tilde g}_{\mu \nu}=g_{\mu \nu} + N^\mu_{in} N^\nu_{out}+N^\mu_{out} N^\nu_{in}$ is the induced metric. We obtain
\be
\theta_{in}=-{d-1 \over r} \, , \hspace{1cm} \theta_{out}={d-1 \over 2}\left(r-{m(v)\over r^{d-1}} \right)\, ,
\ee
and hence the apparent horizon is located at
\be
r_{h}=(m(v))^{1/d} \, .
\label{apph}
\ee  
We will also call apparent horizon the co-dimension one surface defined by the union of the surfaces \eqref{apph}.

\begin{figure}[thbp]
\centering
\includegraphics[width=7cm]{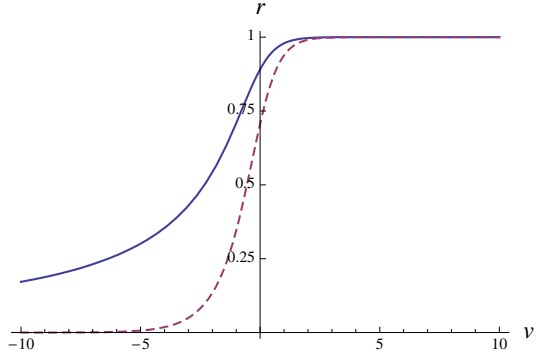} 
\caption{\label{fig:horizon} Position of the event (solid) and the apparent (dashed) horizons of the Vaidya metric \eqref{vaidya} for $d\!=\!2$ and $m(v)\!=\!(\tanh v\!+\!1)/2$.}
\end{figure}

Satisfying the null-energy condition \cite{Hawking:1973uf}, $T_{\mu \nu} N^\mu N^\nu \geq 0$ for any null vector $N^\mu$, implies
\be
{d m \over dv}\geq0 \, .
\label{nec}
\ee
Under this condition, the apparent horizon always grows. We will assume that the mass function $m(v)$ tends to a finite value, $m$, as $v\!\to\!\infty$. The location of the event horizon is given by
 \be
 {dr_e \over dv}={1 \over 2} \left(r^2-{m(v) \over r^{d-2}}\right) \, , \hspace{1cm} \lim_{v\to \infty}r_e(v)=m^{1/d}\, .
 \ee
It always lies at larger values of the radial coordinate than the apparent horizon, see Fig.\ref{fig:horizon}. Contrary to the apparent horizon, the event horizon is different from zero well before the collapse process starts.

When the mass function is constant, $m(v)\!=\!m$, \eqref{vaidya} reproduces the Schwarzschild back hole geometry with planar horizon
\be
ds^2=-\left( r^2-{m\over r^{d-2}} \right) dt^2+{dr^2 \over r^2-{m\over r^{d-2}}} +r^2 \sum_{i=1}^{d-1} dx_i^2 \, ,
\label{schwarz}
\ee
as can be seen by the coordinate change 
\be
v=t+f(r) \, , \hspace{1cm} {df \over dr} = {r^{d-2} \over r^d -m} \, .
\label{tvchange}
\ee
The relation between the coordinates $v$ and $t$ in the  \eqref{schwarz} geometry, close to the boundary and the event horizon is
\begin{eqnarray}
v & \simeq & t-{1\over r}  \, , \hspace{4.2cm} r\rightarrow \infty  \nonumber\\
v & \simeq & t+{1 \over d m^{1/d}} \,{\rm ln} \big( r-m^{1/d} \big) \, , \hspace{1cm}  r\rightarrow m^{1/d}
\label{vt}
\end{eqnarray}
Close to the boundary the foliations defined by both coordinates asymptotically coincide and define the same boundary time.
However they differ strongly in the interior. In particular close to the event horizon $t$ finite implies $v\!\rightarrow \!-\infty$ and conversely $v$ finite $t\!\rightarrow\! \infty$.

In the following we will focus on the AdS$_3$/CFT$_2$ duality. For static backgrounds, like AdS$_3$ and the BTZ black hole with non-compact horizon described by \eqref{schwarz} with $d\!=\!2$, the holographic evaluation of the entanglement entropy requires to study spacelike geodesics that live in a constant $t$ slice \cite{Ryu:2006bv,Ryu:2006ef}. The central segment of the geodesic will probe smaller values of the radial coordinate as the distance between its endpoints, anchored at the AdS boundary, is increased. Since $v$ is a monotonously increasing function of $r$ outside the event horizon, the geodesics will at the same time access earlier and earlier values of $v$. Moreover, those long enough to approach the event horizon, using \eqref{vt}, must reach $v\!\to\!-\infty$. We will show in the next section that geodesics evaluated in the Vaidya geometry have a similar behavior in terms of the $v$ coordinate. Namely, for any fixed boundary time they explore arbitrarily early values of $v$. The map that this implies between measurements at a given time in the dual field theory and bulk physics will determine the evolution of the entanglement entropy.

\section{\label{geodesics}Spacelike geodesics in AdS$_3$-Vaidya}

The AdS$_3$-Vaidya metric with boundary $R^{1,1}$ is given by
\be
ds^2=-\big( r^2-m(v) \big) dv^2+2 dr dv+r^2 dx^2 \, .
\label{vaidya3}
\ee
Since we are interested in describing a black hole formation process, $m(v)$ must interpolate between $0$ and a finite value $m$, which sets the final black hole temperature $T\!=\!\sqrt{m}/2\pi$. In this paper we will consider the analytic function
 \be
m(v)=m \, {{\rm tanh} (v/a)+1 \over 2}  \, .
\label{tmass}
\ee
We will denote the boundary value of $v$ by $t$, the field theory time.
This geometry provides the holographic dual of a strongly coupled 2d CFT on a line in its ground state for $t\!\lesssim\!-2a$. Around $t\!=\!0$ the ground state is perturbed in a translationally invariant way creating a finite energy density that smoothly builds up in a time extent $\delta t \! \sim\! 4a$. The system will then evolve towards a state of thermal equilibrium at the temperature set by the black hole collapse process. The expectation value of the field theory energy-momentum tensor can be read from the variation of the dual gravity action, properly regularized, with respect to the boundary metric \cite{Balasubramanian:1999re,deHaro:2000xn}. We obtain for the field theory energy density and pressure 
\be
\epsilon(t)=p(t)=m(t) \, .
\label{energy}
\ee

In \cite{Hubeny:2007xt} it was proposed that the holographic derivation of the entanglement entropy \cite{Ryu:2006bv,Ryu:2006ef} extends to dynamical backgrounds. 
The entanglement entropy for an interval of size $l$ is then given by the length of the  spacelike geodesic, most conveniently parameterized by the coordinate $x\in [0,l]$, and described by functions $r(x)$ and $v(x)$ subject to the boundary conditions
\be
r(0)=r(l)=\infty \, , \hspace{1cm} v(0)=v(l)=t \, .
\label{bc}
\ee
With this parameterization, the length functional for AdS$_3$-Vaidya is
\be   
L=\int_{0}^{l} dx \sqrt{r^2+2 r' v' -(r^2-m(v)) v'^2} \, .
\label{lengthfunc}
\ee
The absence of explicit $x$-dependence in the integrand leads to the following integral of motion
\be
{r^4 \over r_{\ast}^2}=r^2 +2 r' v'-(r^ 2-m(v)) v'^ 2 \, ,
\label{eom1}
\ee
where $r_{\ast}$ is a constant. 
The two second order differential equations obtained from extremizing \eqref{lengthfunc}
can be reduced with the help of the previous conservation equation to
\be
r^ 2-r^2 v'^2-r v'' +2 v' r'=0 \, .
\label{eom2}
\ee
The boundary conditions \eqref{bc}, together with the translational invariance of the metric in the $x$ direction, imply that the associated geodesics will be symmetric with respect to the midpoint, at which $r'\!=\!v'\!=\!0$. For this type of geodesics  the integration constant $r_\ast$ gives the value of $r$ attained at the midpoint. 
The previous differential equations with the mass function \eqref{tmass} can only be integrated numerically. Some results for the case of compact $x$ were already presented in \cite{Hubeny:2007xt}. Without loss of generality, we will perform the numerical integration of the geodesic equations for a final black hole mass $m\!=\!1$, or equivalently $\beta\!=\!2\pi$. The parameter $m$ can be reintroduced by a simple rescaling, under which the time extent of the perturbation transforms as $a\!\to\!a/\sqrt{m}$.

Substituting \eqref{eom1} into the length functional and using the symmetry of the geodesics, we obtain the simplified expression
\be  
L(l,t)={2 \over  r_\ast}   \int_{0}^{l/2} r(x)^2  dx \, .
\label{length}
\ee
The equations \eqref{eom1}, \eqref{eom2} for large values of the radial coordinate reduce to those describing geodesics in AdS$_3$. As in that space, the leading behavior close to the boundary is given by
\be
x\simeq {r_\ast \over 2 r^2} \, , \hspace{1cm} v\simeq t-{1 \over r}+{b \over 2 r^2}   \, ,
\label{larger}
\ee
where we have focused on a neighborhood of the geodesic endpoint $x\!=\!0$. The integration constants $r_\ast$ and $b$ are to be fixed by requiring that the boundary conditions \eqref{bc} are fulfilled at $x\!=\!l$.
This implies $b\!=\!0$ both for AdS$_3$ and the BTZ black hole, while for the dynamical Vaidya geometry we will show below that this is in general not the case. The behavior of $r(x)$ close to the boundary gives rise to a logarithmic divergence of the geodesic length, which can be regularized by cutting the integration in \eqref{length} at $\eta\!>\!0$. According to the holographic dictionary, $\epsilon\!=\!1/r(\eta)$ is to be interpreted as a field theory UV cutoff. This provides the holographic counterpart to the necessity of introducing a UV cutoff for defining the entanglement entropy \cite{Srednicki:1993im}. Hence
\be
S(l,t)={c \over 6}L(l,t)= {c\over 3\, r_\ast}  \int_{\eta}^{l/2} r(x)^2  dx   \, .
\label{eehol}
\ee
The value of the central charge can be read from \eqref{holents}, $c\!=\!3/2 G^{(3)}_N$.

\begin{figure}[thbp]
\centering
\begin{tabular}{cc}
\includegraphics[width=7cm]{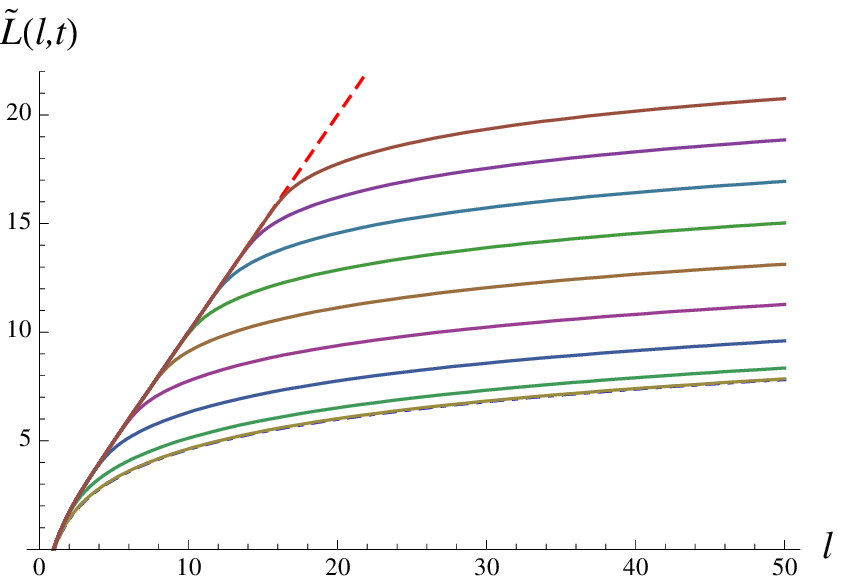} & \includegraphics[width=7cm]{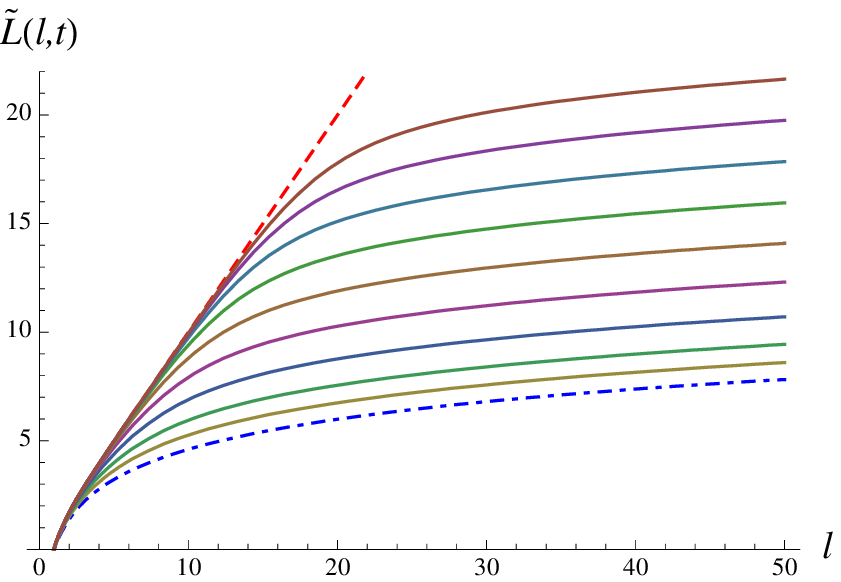}
\end{tabular}
\caption{\label{fig:Sfuncionl} ${\tilde L}(l,t)$ as a function of $l$ for $t=0,1,..,8$ from bottom to top, for $a\!=\!1/3$ (left) and $a\!=\!2$ (right). The dashed line corresponds to thermal equilibrium and the dot-dashed line to the vacuum result.}
\end{figure}

We shall present now the results from the numerical integration of the geodesic equations. For convenience we will plot the finite, cutoff independent contribution to the geodesic length defined by
\be
{\tilde L}(l,t)=  {2\over r_\ast}  \int_{\eta}^{l/2} r(x)^2  dx -2{\rm log}\,r(\eta) \, .
\label{eeind}
\ee
In Fig.\ref{fig:Sfuncionl} we show this quantity as a function of the interval size at fixed time, for several times that cover from the beginning of the perturbation until well after it has ceased. 
The logarithmic dependence proper of the CFT vacuum \eqref{eevac} is reproduced for small intervals, since the associated geodesics stay close to the boundary and perceive an AdS geometry. A regime of linear growth is established at intermediate values of $l$ some time after the action on the system ends. The slope in this regime is proportional to the final temperature characterizing the black hole formation process. However for perturbations giving rise to a slowly varying geometry, as in Fig.\ref{fig:Sfuncionl}b, the geodesic length is able to reflect the building up of $m(v)$. For large intervals there is an asymptotic return to a logarithmic dependence on $l$. The effect of the perturbation is encoded in a time dependent shift
\be
S(l,t) \to s(t) + {c \over 3} \,{\rm log} {l \over \epsilon} \, .
\label{eeasymp}
\ee
The transition from the linear to the logarithmic regime happens at $l \!\simeq \! 2t$, being sharper the shorter the time extent of the perturbation is. This is clearly seen comparing Fig.\ref{fig:Sfuncionl}a and \ref{fig:Sfuncionl}b.

\begin{figure}[thbp]
\centering
\begin{tabular}{cc}
\raisebox{1mm}{\includegraphics[width=7cm]{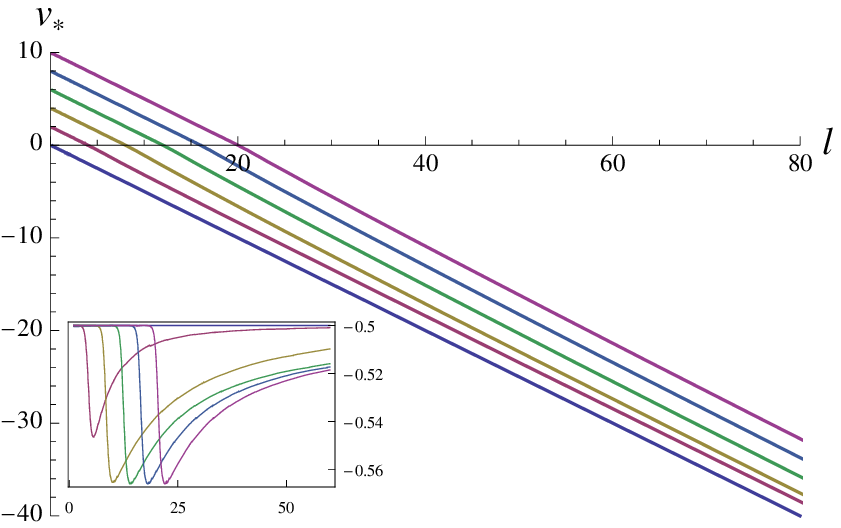}} & \includegraphics[width=7cm]{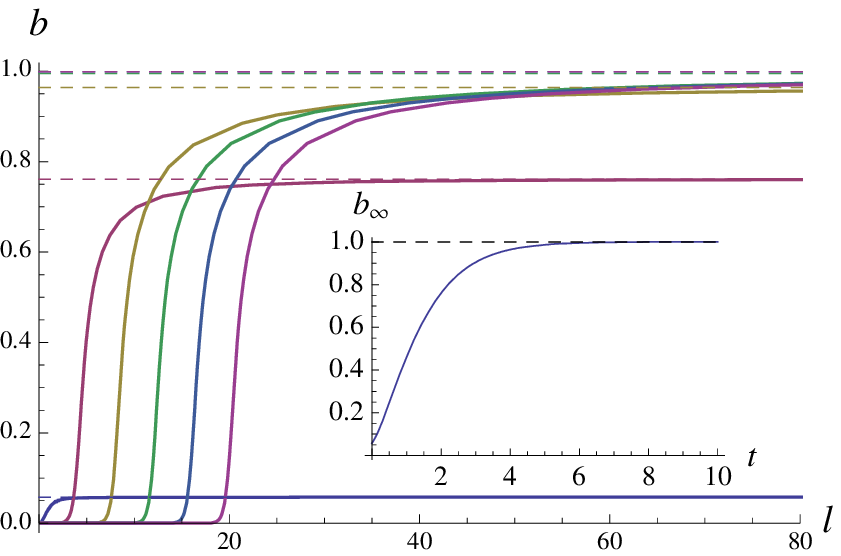}
\end{tabular}
\caption{\label{fig:vsfuncionl} For $a\!=\!1/3$ and $t\!=\!0,2,..,10$: (left) value of $v$ at the geodesic midpoint as a function of the interval size, and on the inset its derivative $\partial_l v_\ast$; (right) $b$ in \eqref{larger} as a function of $l$ (solid lines) and its asymptotic value $b_\infty$ (dashed lines), whose evolution is plotted on the inset.}
\end{figure}

We will denote by $v_\ast$ the value of $v$ attained at the geodesic midpoint. Its dependence on the size of the interval is plotted in Fig.\ref{fig:vsfuncionl}a. As advanced in the previous section, $v_\ast$ decreases as $l$ increases. For the case we are considering of a non-compact $x$ direction, $v_\ast \rightarrow -\infty$ when the interval tends to cover the complete system. Hence, no matter how late a field theory time we consider, the geodesics \eqref{bc} are able to probe the whole collapse process.  

The linear regime of $S(l,t)$ as a function of $l$ can be easily understood. Although the mass function \eqref{tmass} has non-compact support, it reaches $98\%$ of its final value at $v\!\sim\!2a$. 
From then on it is reasonable to consider the black hole as already formed and described by a BTZ solution to a very good approximation. BTZ geodesics with the boundary condition \eqref{bc} satisfy the very simple relation $l\!=\!2(t-v_\ast)$.\footnote{This can be deduced from
\be
v=t-{1 \over 2 \sqrt{m}} \, {\rm ln} \,{r + \sqrt{m}\over r -\sqrt{m}} \, , \hspace{1cm} 
l= {1 \over \sqrt{m}} \,{\rm ln} \,{r_\ast + \sqrt{m}\over r_\ast -\sqrt{m}} \, .
\ee 
The fist equation is derived from \eqref{tvchange}, and substituting it into the integral of motion \eqref{eom1} with $m(v)\!=\!m$ the second is obtained after integration.}
Therefore the condition for geodesics to lie entirely in the BTZ part of the geometry is 
\be 
l< 2t-4a \, .
\label{linearee}
\ee 
Those that reach the black hole horizon extend tangent to it, and induce a linear dependence of $S(l,t)$ with the interval size. The BTZ geodesics reproduce the result for the entanglement entropy at thermal equilibrium \eqref{eethermal} \cite{Ryu:2006bv,Ryu:2006ef}. From that expression we observe that the linear behavior requires intervals of size $l\!\gtrsim\!\beta/2$, where $\beta\!=\!2\pi/\sqrt{m}$ is the inverse temperature. For them
\be
S_\beta(l)={c \over 3} \log {\beta \over 2 \pi \epsilon}+ {\pi c l \over 3 \beta} \, .
\label{extee}
\ee
Compatibility with the bound \eqref{linearee} implies that a clear extensive regime for the cutoff independent piece of the entanglement entropy is not established before
\be
t_0\simeq \beta/4+2a \, .
\label{t0}
\ee
The influence of the parameter $a$ on $t_0$ can be checked in Fig.\ref{fig:Sfuncionl}. 
 
\begin{figure}[thp]
\begin{center}
\begin{tabular}{lc}
    \subfigure[]{\label{fig:geo-a}\includegraphics[scale=0.7]{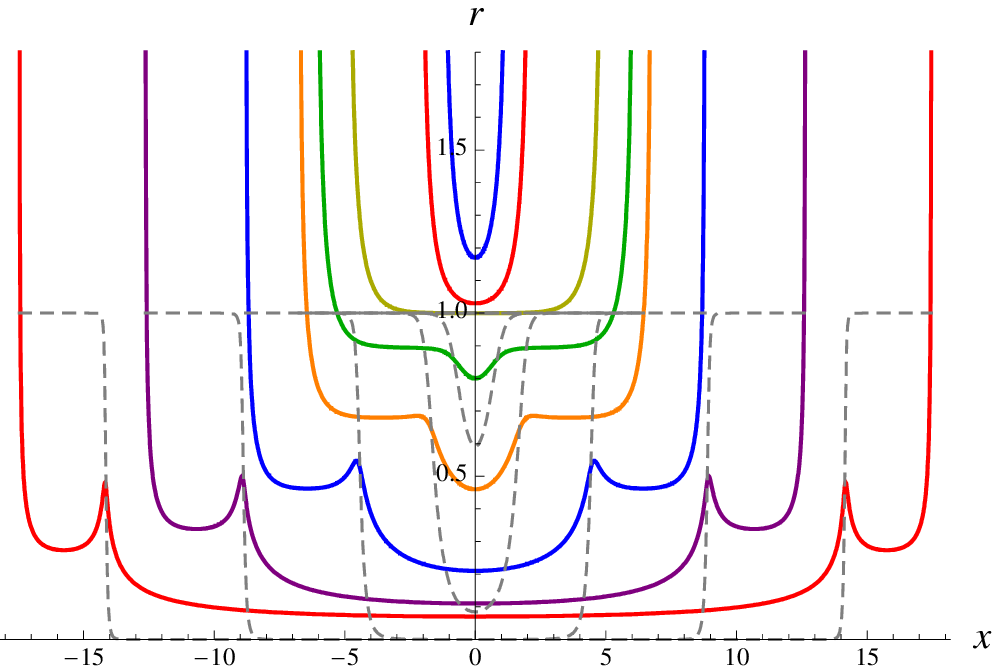}} 
    \subfigure[]{\label{fig:geo-b}\includegraphics[scale=0.7]{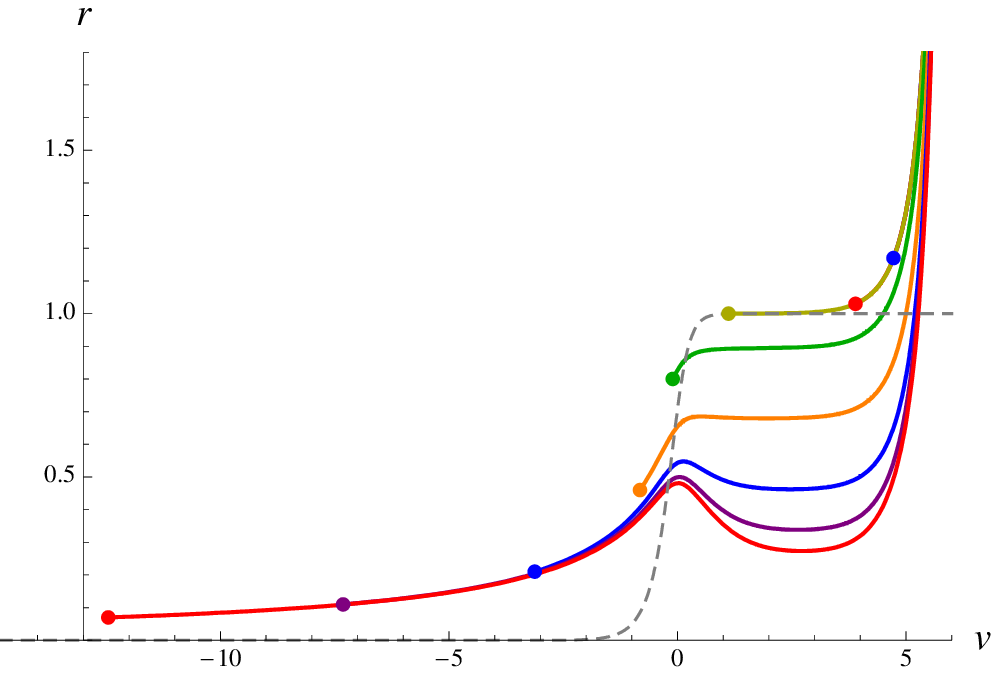}}  \\
    \subfigure[]{\raisebox{1cm}{\label{fig:geo-c}\includegraphics[scale=0.7]{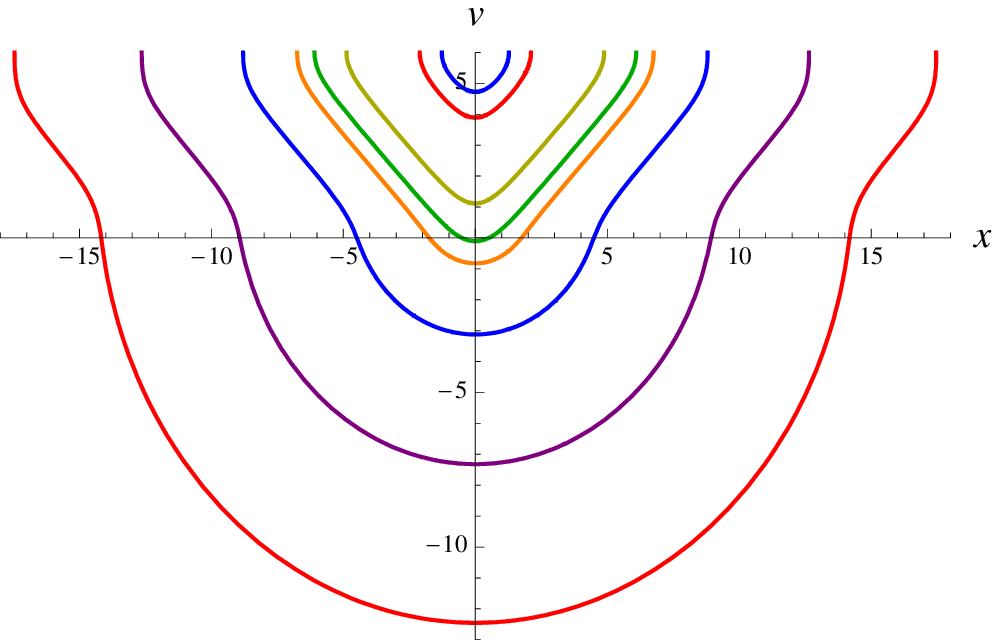}}} 
  \hspace{0.75cm}\subfigure[]{\label{fig:geo-d}\includegraphics[scale=0.28]{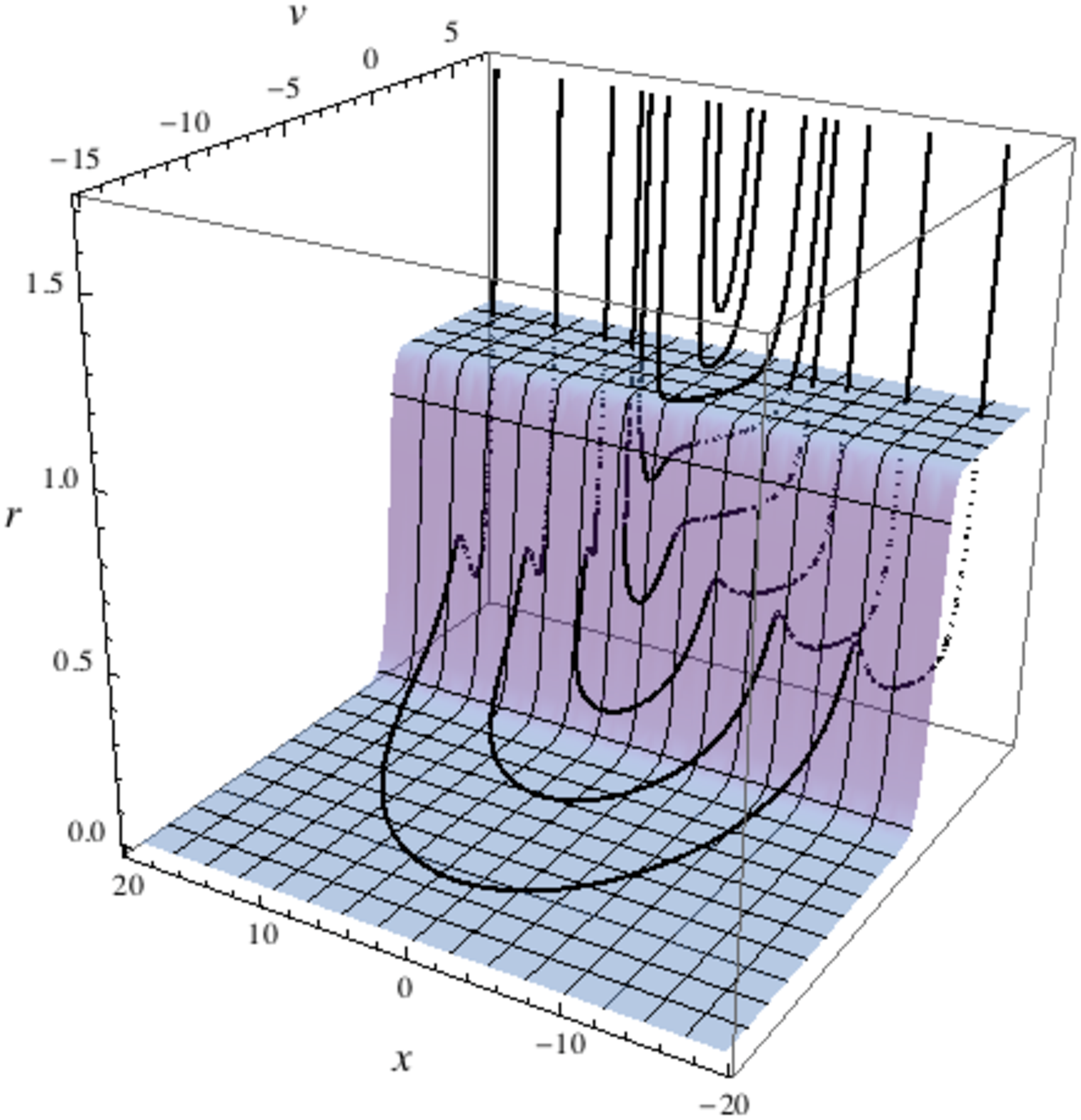}} 
\end{tabular}
\end{center}
\caption{\label{fig:geo} Profile of geodesics for different $l$ at $t\!=\!6$ and $a\!=\!1/3$. Projection into the (a) $r\!-\!x$ plane, (b) $r\!-\!v$ plane and (c) $v\!-\!x$ plane. (d) Full profile of the geodesics. The dashed line in (b) and the shaded surface in (d) signal the location of the apparent horizon. In (a) the dashed lines represent the position of the apparent horizon at the value of $v$ defined by the trajectory of each geodesic: $\sqrt{m(v(x))}$.}
\end{figure}
 
Geodesics associated to intervals larger than \eqref{linearee} enter the black hole formation region and present qualitative new features. In this region the apparent horizon builds up and clearly differs from the event horizon. Let us describe first the behavior of geodesics with endpoints at $t\!>\!t_0$. When their central segment starts feeling the varying geometry, it develops a dip that however does not reach the apparent horizon. This region is followed by a segment that crosses behind the apparent horizon and extends at almost constant $r$ into the BTZ region, see Fig.\ref{fig:geo}. Both segments lie behind the event horizon. This behavior becomes more pronounced as the size of the interval increases. Once $v_\ast$ reaches the pure AdS part of the geometry, the dip around the midpoint broadens. The transition to large values of the radial coordinate is now through a region well behind the apparent horizon, where a local minimum in $r$ develops. It is interesting to point out that geodesics that enter the black hole formation region require the integration constant $b$ in \eqref{larger} to be different from zero, as shown in Fig.\ref{fig:vsfuncionl}b.\footnote{The dependence of the bound \eqref{linearee} on the parameter $a$ can be neatly seen in this figure, since $b$ starts differing from zero before $2t$.} Hence, although a segment of the geodesic lives in the BTZ part of the geometry, it differs from that of an equal time BTZ geodesic. This is the technical reason that allows these geodesics to cross behind the horizon already in the BTZ region.

\begin{figure}[thp]
\centering
\begin{tabular}{cc}
\includegraphics[width=7cm]{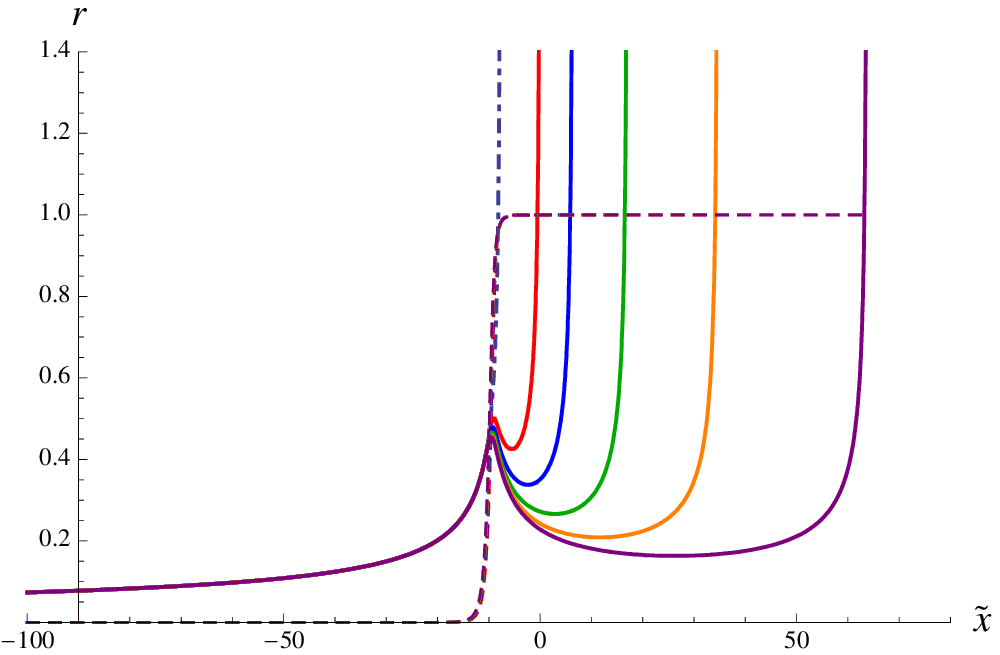} & \hspace{5mm}\raisebox{-10mm}{\includegraphics[width=5.5cm]{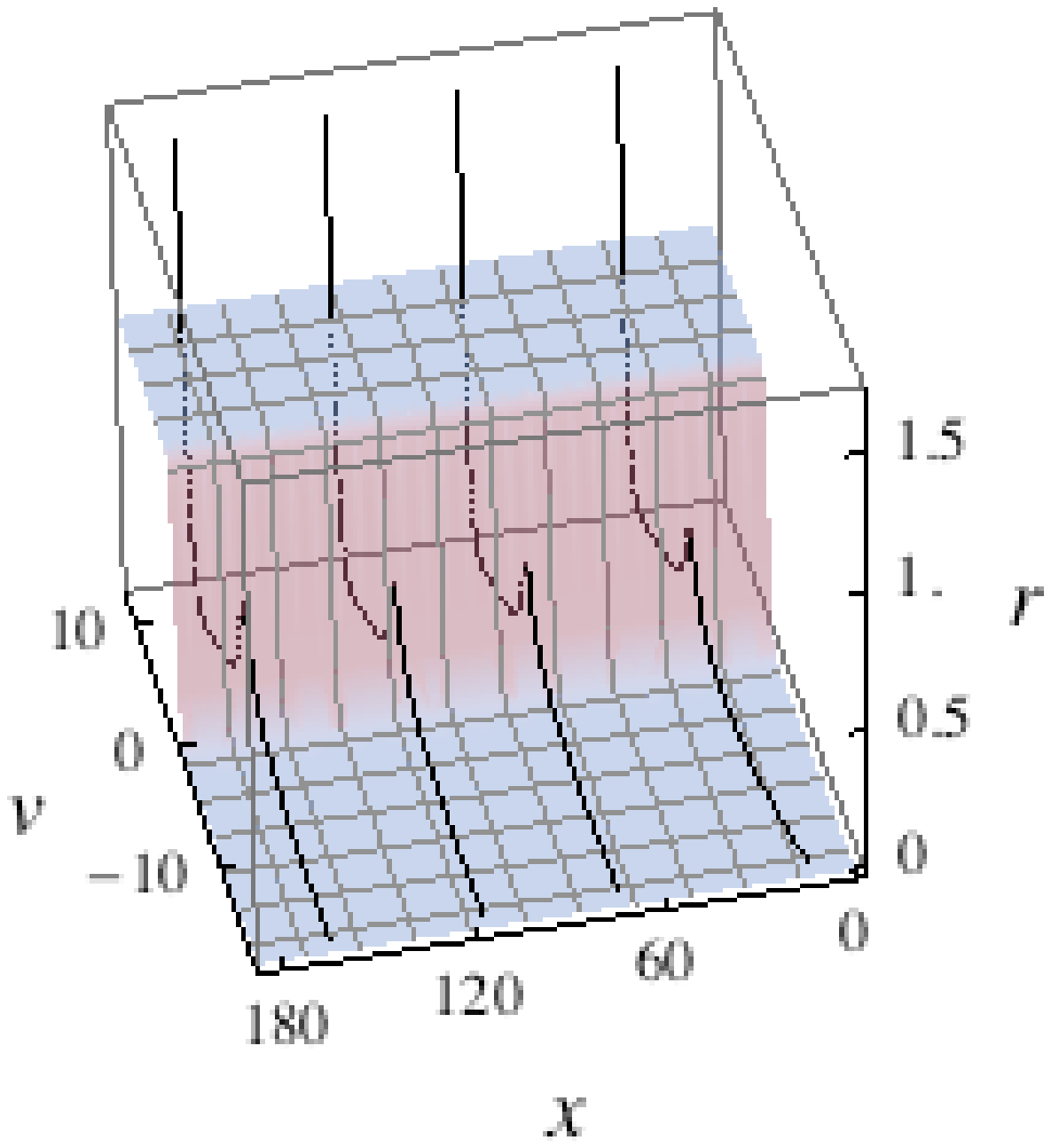}}
\end{tabular} 
\caption{\label{fig:geoinfrx} (left) Radial projection of the infinite interval geodesics as a function of the rescaled variable ${\tilde x}$ for $a\!=\!1/3$ and $t=3,3.5,4,4.5,5$ from left to right. The dashed line represents the position of the apparent horizon as defined in Fig.\ref{fig:geo}a, which practically coincide for all these geodesics. The dot-dashed line corresponds to the infinite AdS$_3$ geodesic. (right) Full profile of several $t\!=\!6$ long geodesics.}
\end{figure}

For asymptotically large intervals the AdS segment tends to cover the whole extent of the geodesic in the $x$ coordinate, explaining the return to a logarithmic growth of the entanglement entropy.
The piece behind the apparent horizon lies then mostly in the $(r,v)$-plane as can be seen in Fig.\ref{fig:geoinfrx}b. Indeed, the projection of this piece in the $x$-coordinates localizes at $x\!=\!0,l$ with extent proportional to $r_\ast$. This is precisely what is needed for this segment to contribute a finite amount to the length functional \eqref{length} in the limit $r_\ast\!\to\!0$, and account for the function $s(t)$ in \eqref{eeasymp}. The limiting large $l$ behavior can be best studied by rescaling $x\rightarrow {\tilde x}\!=\!x/r_\ast$ and setting $r_\ast=0$ in equations \eqref{eom1}, \eqref{eom2}
\be
r^4= 2 r' v'-(r^ 2-m(v)) v'^ 2 \,  , \hspace{1cm} 2 v' r'-r^2 v'^2-r v'' =0 \, .
\label{eominf}
\ee
These equations should be integrated with the boundary conditions $r(0)\!=\!\infty$, $v(0)\!=\!t$. The remaining integration constant is fixed by requiring that the solution reproduces a fixed time geodesic in AdS$_3$ far from the boundary, see Fig.\ref{fig:geoinfrx}a. It is natural to identify the length of this AdS$_3$ geodesic with the $l\!\to\!\infty$ limit of the logarithmic term in \eqref{eeasymp}. The function $s(t)$ can then be represented in terms of the length difference between the Vaidya and the AdS$_3$ limiting geodesics
\be
s(t)= {c \over 6} \Delta L_\infty(t)= {c \over 3}  \left[  \int_{\eta}^\xi r({\tilde x})^2  d{\tilde x} -{\rm ln}\,{r(\eta) \over r(\xi)} \right] \, .
\label{sent}
\ee
The point $\xi$ has been chosen in the region where both geodesics coincide. The AdS$_3$ limiting geodesic is described by
\be
{\tilde x}= {\tilde x}_0+{1\over 2 r^2} \, , \hspace{1cm} v= t-{1 \over r} \, ,
\ee
where the matching to the Vaidya geodesic far from the boundary determines the endpoint ${\tilde x}_0$. The second term of the bracket in \eqref{sent} gives the regularized length of the AdS$_3$ geodesic segment between $\xi$ and the boundary. 

\begin{figure}[thp]
\centering
\begin{tabular}{cc}
\includegraphics[width=7cm]{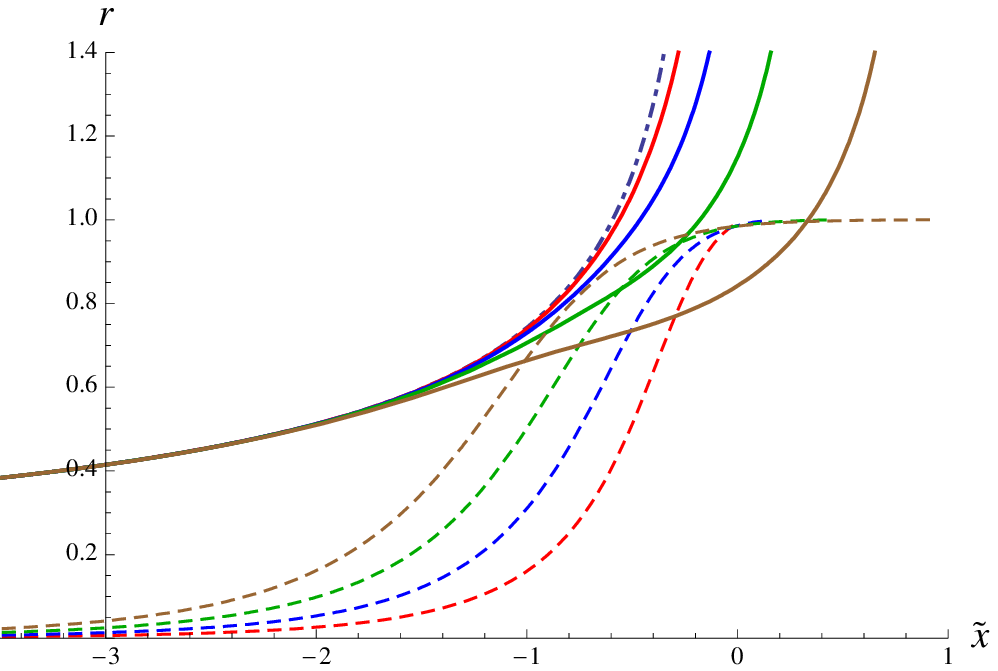} & \includegraphics[width=7cm]{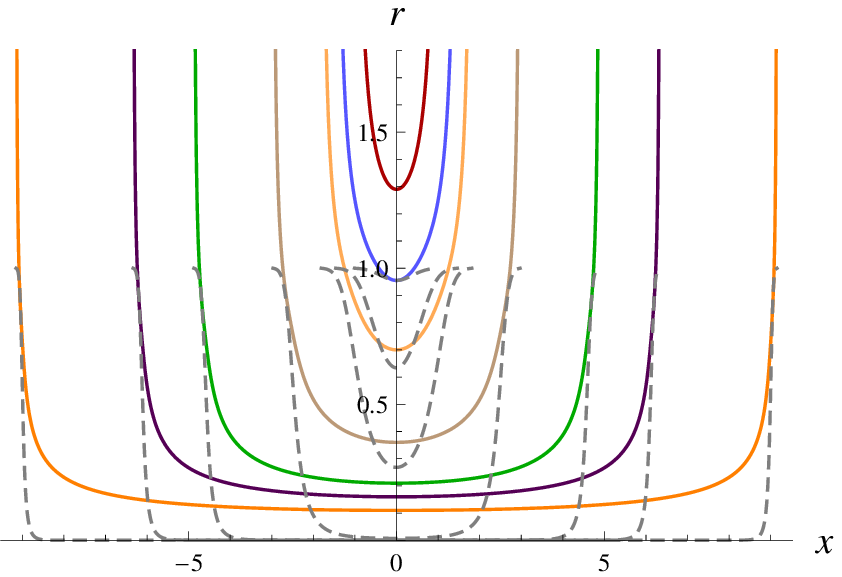}
\end{tabular} 
\caption{\label{fig:geoearly} (left) Infinite interval geodesics at $t\!=\!0.8,1,1.4,1.7$ and $a\!=\!1/3$
from left to right and the associated projection of the apparent horizon as in Fig.\ref{fig:geo}a from right to left (dashed lines). The dot-dashed line corresponds to the infinite AdS$_3$ geodesic. (right) Projection in the $r-x$ plane of several geodesics for $t\!=\!1.4$.}
\end{figure}

To complete our analysis, we consider the early time behavior of the geodesics. Contrary to what happens for $t\!>\!t_0$, at early times the geodesics enter the black hole formation region without necessarily reaching the apparent horizon. Moreover there is a threshold time before which geodesics stay always outside it. This happens after the onset of the perturbation. Taking as an example $\beta\!=\!2\pi$ and $a\!=\!1/3$, from Fig.\ref{fig:geoearly} we observe that geodesics of any length stop crossing the apparent horizon before $t\!\simeq\!1.4$, well after the perturbation has ended. At that time $S(l,t)$ already differs appreciably from the vacuum result, with $s(t)$ being of order $c/6$ as can be seen in Fig.\ref{fig:Sfuncionl}a. 

Although the apparent horizon is not a necessary ingredient in the holographic derivation of the entanglement entropy, its role in it could indicate how much information it encodes about the entanglement properties of the field theory. An important related question is wether the apparent horizon can be of relevance to define a notion of field theory entropy in far from equilibrium situations \cite{Hubeny:2007xt,Figueras:2009iu}. Contrary to the case of the event horizon in static backgrounds, we have seen that the apparent horizon does not act as a wall for the geodesics involved in the calculation of the entanglement entropy. However for $t\!>\!t_0$ it strongly influences the geodesic shape and length, with $s(t)$ measuring the geodesic segment behind it. Not even this applies in the early, far from equilibrium stages of the evolution, when the apparent horizon does not seem to play a role in what regards the holographic derivation of the entanglement entropy.

\section{\label{comparison}Propagation of entanglement} 

The results of the previous section show a strong analogy between our holographic model and the critical quantum quenches studied in \cite{Calabrese:2005in}. In both cases the cutoff independent piece of the entanglement entropy behaves after a certain time as an extensive quantity up to intervals of size $l\!\simeq\!2t$, see equation \eqref{eeevol} and Fig.\ref{fig:Sfuncionl}. Differences arise however for larger intervals. For them $S(l,t)$ turns out to reproduce the entanglement properties of the initial sea of the excitations generated by the perturbation that brings the system out of equilibrium. The excited state created by a quantum quench has a mass gap. Therefore only excitations produced at close points can be entangled, and in that case $S(l,t)$ saturates to a constant value for $l\!\gtrsim\!2t$. The Vaidya geometry models a perturbation on the CFT vacuum which acts during a time $\delta t \!\simeq\!4a$. Remarkably the quantum entanglement among the resulting excitations appears to be characterized by the same logarithmic scaling as in the CFT vacuum. Indeed at $t\!=\!0$, when the perturbation is half way of being completed,  $S(l,0)\!\simeq \! {c\over 3} \,{\rm log}\, l$ up to a subleading term of order ${\cal O}(a^{2})$. This implies the presence of long range correlations among the excitations sourced by the perturbation, contrary to a quantum quench. However as in that case we have seen that for all times the entanglement entropy of large intervals reproduces the initial entanglement pattern, which now implies the return to a logarithmic growth. 

The preservation of the logarithmic scaling of the entanglement entropy during the perturbation, and afterwards for large intervals, suggests that the complete evolution in the holographic equilibration process is unitary, as it happens in a quantum quench. Since the field theory starts in a pure state, a unitary evolution will preserve this character. We can use the property that in a pure state the entanglement entropy of a region is equal of that of its complementary, as stated in \eqref{aabar}, to study this question. According to \cite{Fursaev:2006ih}, the geodesic whose length should give the entanglement entropy of a certain region does not need to be connected. It is only required that this geodesic, together with the region under consideration, defines the boundary of a certain surface in the bulk geometry. For the infinite, complementary region to a finite interval we have clearly two choices \cite{Headrick:2010zt}. A single geodesic with the same boundary conditions \eqref{bc} but extending towards $x\!\to \!\pm\infty$. The second possibility is composed of two disjoint pieces: the finite size geodesic whose length gives $S(l,t)$, plus a solution of the equations of motion that covers $x\!\in\!(-\infty,\infty)$ without reaching the AdS boundary. Such a solution is provided by $r\!=\!0$ at $v\!=\!-\infty$, where the mass function \eqref{tmass} strictly vanishes and hence there is no curvature singularity at the origin of the radial coordinate. Substituting in \eqref{lengthfunc} we obtain that it has zero length, as it would be the case in empty AdS. Hence the disconnected choice of geodesics gives the minimal length and is the preferred one. It implies that  the entanglement entropy of a finite interval equals that of its complementary region for any time, proving that the evolution is unitary.\footnote{In the previous section we have seen that geodesics associated to infinite size intervals anchoring in the AdS boundary at arbitrary finite time reach $v\!=\!-\infty$, as shown in Fig.\ref{fig:vsfuncionl}a. The same happens in static backgrounds. This offers additional support for the previous construction.}
 
\begin{figure}[thp]
\centering
\begin{tabular}{cc}
\includegraphics[width=7cm]{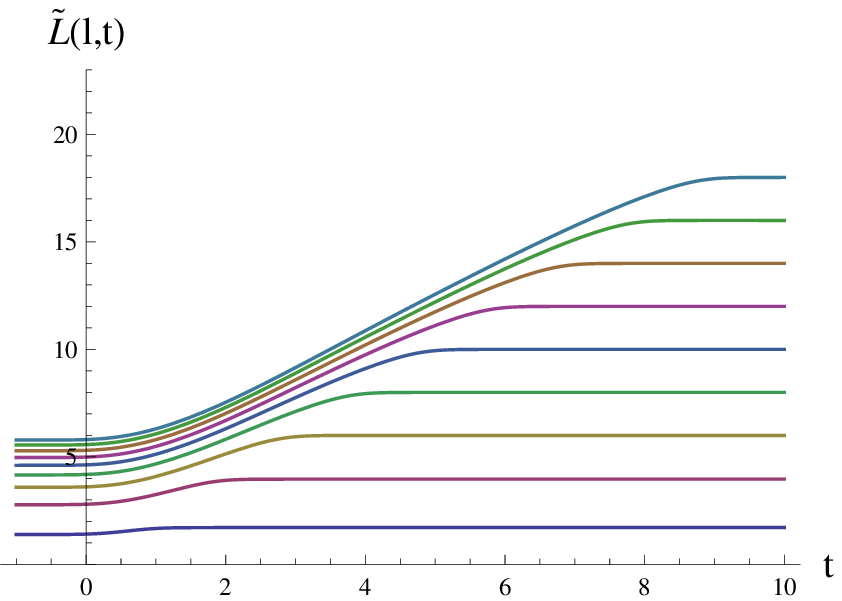} & \includegraphics[width=7cm]{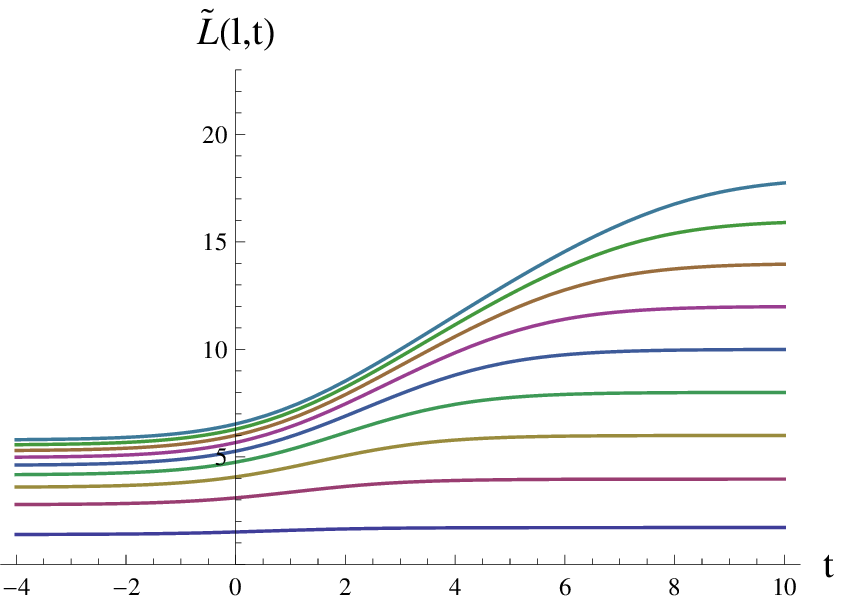}
\end{tabular}
\caption{\label{fig:Lfunciont} ${\tilde L}(l,t)$ as a function of $t$ for $l\!=\!2,..,18$, for $a\!=\!1/3$ (left) and $a\!=\!2$ (right) from bottom to top.}
\end{figure}

The evolution of the entanglement entropy for several fixed intervals is plotted in Fig.\ref{fig:Lfunciont}. We observe that $S(l,t)$ for a given $l$ attains its final value within a finite time. The holographic explanation for this fact is that after a finite time the associated geodesics lie in the BTZ part of the geometry and thus produce a time-independent result. An estimation for the time at which this happens is  obtained from \eqref{linearee} 
\be
t\simeq {l \over 2} +2a \, .
\ee
Except for the dependence on the parameter $a$, whose effect is clearly noticed in Fig.\ref{fig:Lfunciont}, the same result holds for critical quantum quenches. According to the picture in Fig.\ref{fig:prop}, both the extensive behavior of the entanglement entropy for $l\!\lesssim\!2t$ and its saturation as a function of time for $t\!\gtrsim \!l/2$ lead to the important conclusion that entanglement propagates in 2-dimensional CFT's with velocity $v^2\!=\!1$ \cite{Calabrese:2005in}. 
The Vaidya model allows to check this property in an equilibration process with initial long range entanglement. 

\begin{figure}[thp]
\centering
\begin{tabular}{cc}
\includegraphics[width=7cm]{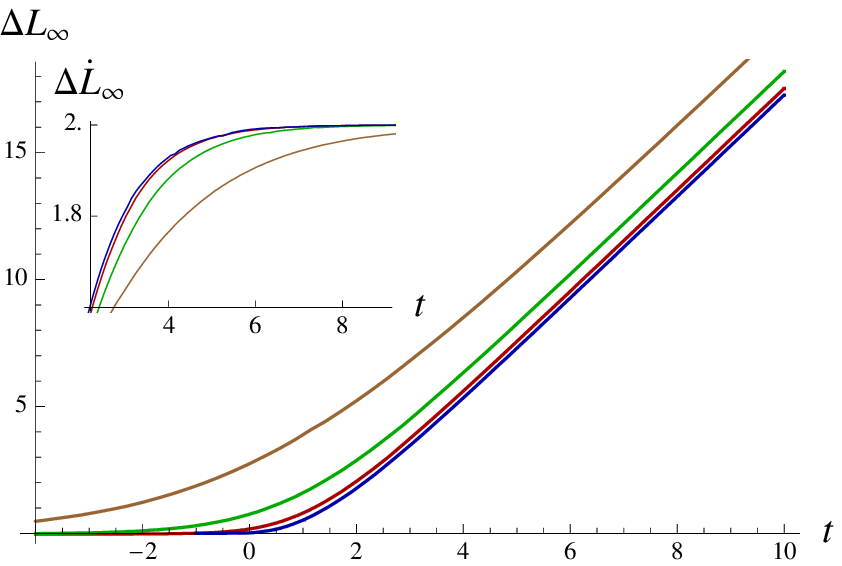} & \raisebox{1mm}{\includegraphics[width=7.5cm]{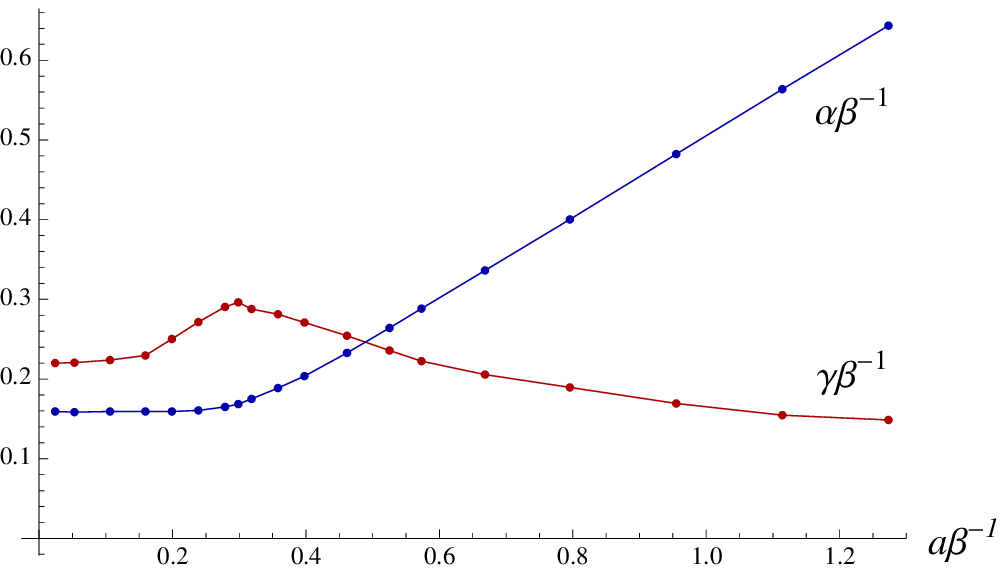}}
\end{tabular}
\caption{\label{fig:geoinf} (left) $\Delta L_\infty (t)$ in \eqref{sent} for $a\!=\!1/3,1,2,4$, and on the inset its time derivative. As in previous figures, $\beta\!=\!2\pi$. (right) The dimensionless quotients $\alpha \beta^{-1}$ and $\gamma \beta^{-1}$ as a function of $a \beta^{-1}$.}
\end{figure}

It is also interesting to analyze the time evolution of the function $s(t)$.  
Equation \eqref{sent} allows its precise numerical evaluation for any time. In Fig.\ref{fig:geoinf}a we have plotted $s(t)$ for several perturbations with different time extent. Its evolution is dominated by a linear growth with twice the slope characterizing the extensive regime \eqref{extee} of the entanglement entropy. The linear behavior and the approach to it are well described by  
\be
s(t) = {\pi c  \over 3 \beta}\left( 2t  + \alpha  \, e^{\gamma-t \over \alpha}\right) -s_0 \, ,
\label{sth}
\ee
where $s_0$, $\alpha$ and $\gamma$ are constants. 

\begin{figure}[thp]
  \begin{center}
  \begin{tabular}{cc}
    \includegraphics[scale=0.37]{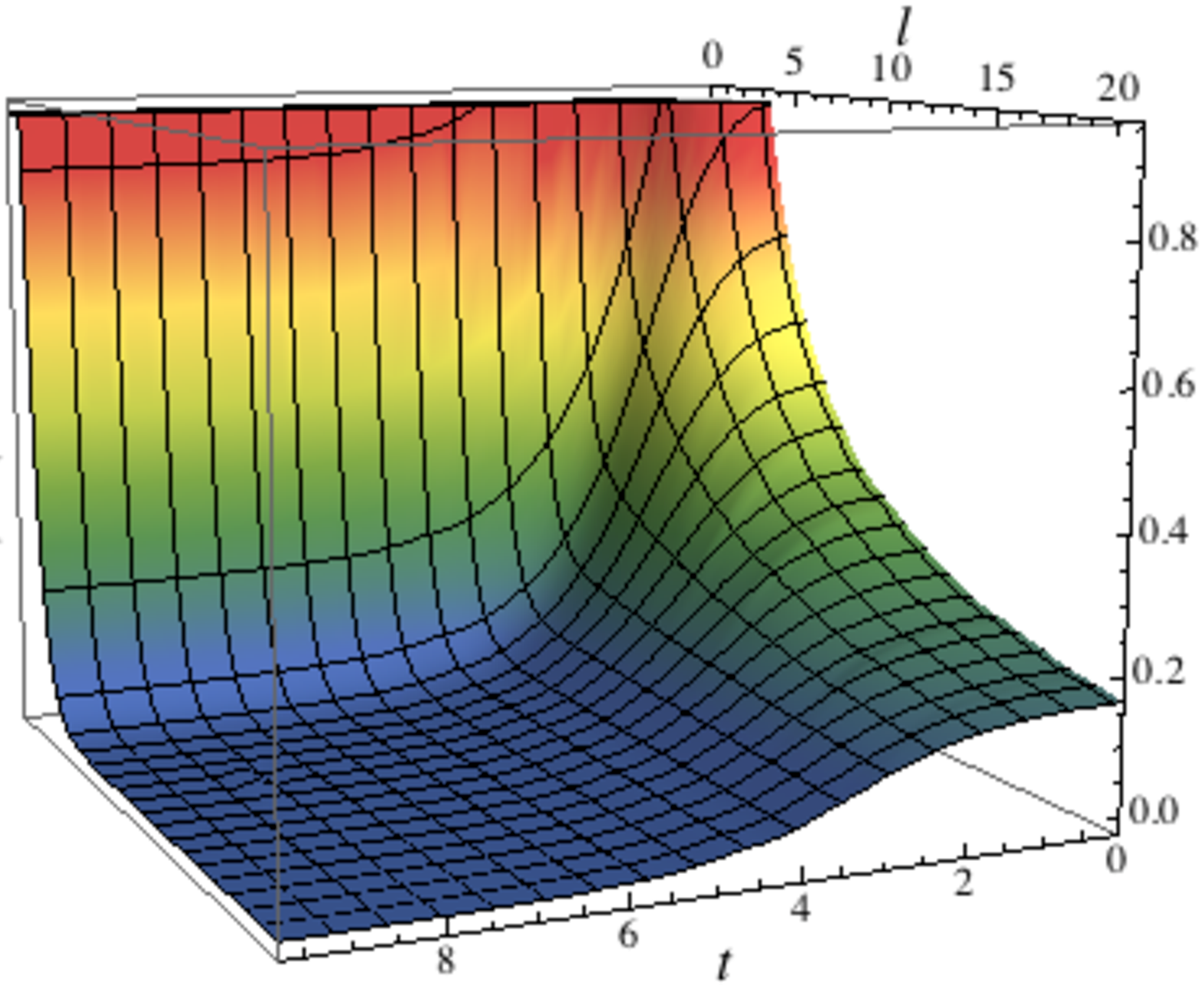}
    \includegraphics[scale=0.35]{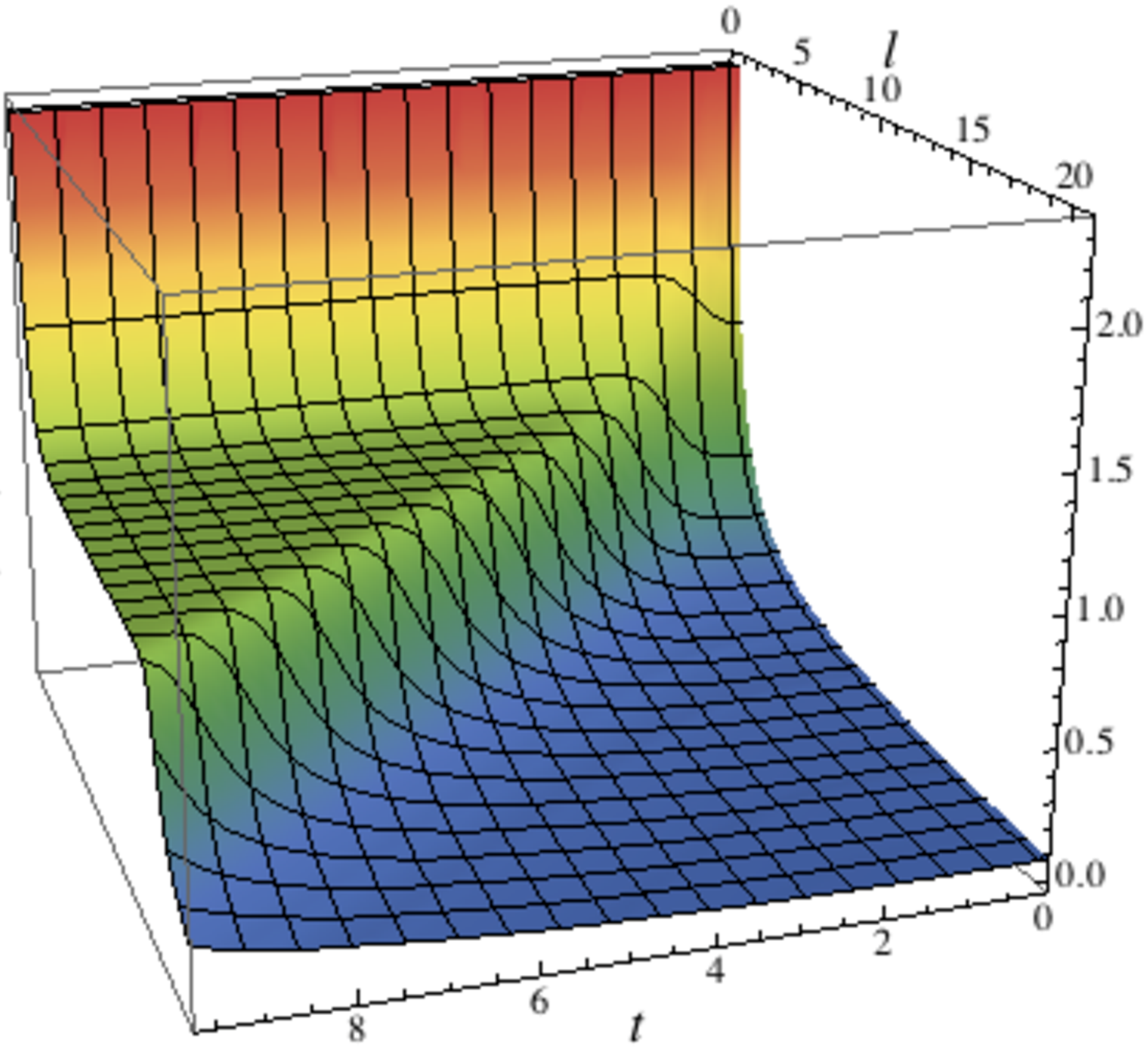}
    \end{tabular}
   \end{center}
  \caption{\label{fig:derivadasL} Plot of $(2\partial_l + \partial_t) (L-\Delta L_\infty)$ (left) and $\partial_l L$ for $a=1/3$ and $\beta\!=\!2\pi$. The difference between $\bar t$ and $t_0$ can be clearly appreciated.}
\end{figure}

The dependence of $\alpha$ and $\gamma$ on the time extent of the perturbation has been plotted in Fig.\ref{fig:geoinf}b. When $a\!\lesssim\!0.2 \, \beta$, both $\alpha$ and $\gamma$ are independent of $a$, implying the same property for the time needed to achieve the linear growth of $s(t)$. Requiring the term in brackets on the rhs of \eqref{sth} to behave linearly within 1$\%$ error leads to a threshold time ${\bar t}\!=\!\gamma\!+\!4\alpha$. For short perturbations we obtain
\be
{\bar t} \simeq 0.85 \, \beta \, ,
\label{tth}
\ee
The fact that \eqref{tth} is independent of the time extent of the perturbation suggests that it has a meaning as a characteristic time scale for the evolution of the holographic plasma. Notice that $\bar t$ occurs latter than $t_0$, the time at which the cutoff independent piece of the entanglement entropy starts exhibiting an extensive regime. In the next section we will discuss a possible interpretation for \eqref{tth}. For long perturbations $a\!\gtrsim\!0.5 \, \beta$, we have instead $\alpha\!=\!a/2$ while $\gamma$ tends to a constant. Hence in this case the threshold time for the linear growth of $s(t)$ is governed by the time extent of the perturbation, and ${\bar t}\!\simeq\!t_0$. 

After ${\bar t}$ not only the asymptotic quantity $s(t)$, but the entanglement entropy itself simplifies. 
Using \eqref{eeasymp}, the following relation is always fulfilled for asymptotically large intervals
\be
(2\partial_l+\partial_t) (S(l,t)- s(t))=0 \, .
\ee
For $t\!>\!{\bar t}$ this equation extends its region of validity to intervals as small as $\beta/2$, the threshold for the onset of the extensive regime, as can be seen in Fig.\ref{fig:derivadasL}a. Integrating we obtain
\be
S(l,t)=f(l-2t)+s(t) \, , \hspace{1cm} t>{\bar t} \, , \,  l>\beta/2 \, ,
\label{eeth}
\ee
which applies to the interesting region $l\!\simeq\!2t$ where the entanglement entropy changes from the extensive to the logarithmic behavior. It implies that after $\bar t$ a regime of steady evolution has been achieved, in which the only change of $S(l,t)$ around $l\!\simeq\!2t$ is a linear increase with time described by the function $s(t)$.
Reproducing the extensive regime of the entanglement entropy \eqref{extee} determines
\be
f(x)={c \over 3} \log {\beta \over 2 \pi \epsilon}+ {\pi c x \over 3 \beta}+s_0 \, , \hspace{1cm} x<-4a \, ,
\label{fmin}
\ee
At large positive values of $x$ the function $f(x)$ inherits from \eqref{eeasymp} a logarithmic behavior
\be
f(x)\simeq{c \over 3}  \log {x+x_0 \over \epsilon} \, .
\label{flarge}
\ee
As an example, for $\beta\!=\!2\pi$ and $a\!=\!1/3$ we have obtained $x_0\!\approx\!7$ which provides a good description already for $x\!\gtrsim\!15$.

Equations \eqref{sth} and \eqref{eeth} also apply to the evolution after a quantum quench \eqref{eeevol}, with the only difference that $f(x)\!=\!0$ for $x\!>\!0$ while \eqref{fmin} holds for $x\!<\!0$. The analysis of quantum quenches performed in \cite{Calabrese:2005in} requires $t,l\!\gg\!\tau_0\!\simeq\!\beta_{eff}$ and hence is valid for times well after the analogue of $\bar t$ and intervals over the threshold for extensive behavior. This is consistent with the regime of validity of \eqref{sth}. In this way we found not only a qualitative but also quantitative agreement between the evolution of entanglement entropy in critical quantum quenches and in our holographic model. The fact that they correspond to quite different type of perturbations strongly supports that a very generic behavior for the entanglement entropy of 2-dimensional CFT's holds even in time dependent situations.

\section{\label{discussion}Comments on thermalization} 

A very interesting aspect of the AdS/CFT correspondence is that it allows to model perturbations which bring the boundary theory far from equilibrium. As for quantum quenches, this process and the subsequent evolution are unitary. Hence if the system starts in a pure state it will remain so for all times, implying that information cannot be lost at the microscopical level. However information will become ever more difficult to access as a result of the propagation and interaction among the excitations sourced by the initial perturbation. This process will drive the system towards a state that at the macroscopical level cannot be distinguished from that of thermal equilibrium. In this section we want to further discuss what can be learnt about the equilibration process and the holographic dictionary implied, from the study of the entanglement entropy. 

When a system is in a pure state, the entanglement entropy gives a measure of the quantum entanglement between the degrees of freedom inside and outside a chosen subsystem \cite{Holzhey:1994we,Vidal:2002rm,Latorre:2003kg,Calabrese:2004eu}. In the case of a dynamical situation, it describes also its evolution. Propagation will make excitations initially localized in a small region spread over ever larger scales. Hence the entanglement entropy of a subsystem will strongly depend on the relation between its size and the typical separation of entangled degrees of freedom at a given time. This effect is very neatly seen in 2-dimensional CFT's, where the study of quantum quenches indicates that this separation evolves as a light front \cite{Calabrese:2005in}. This was also confirmed by our holographic model, whose main difference with a quantum quench is the presence of long range entanglement among the initial sea of excitations. 

Interaction is responsible for the redistribution of energy between the different momentum modes, and therefore it is crucial for the system to relax towards a state of thermal equilibrium. Let us consider the entanglement entropy of a 2-dimensional CFT at thermal equilibrium \eqref{eethermal} and perform its Fourier transform.
We obtain
\be
2\int_0^\infty \cos(pl) S_\beta(l) {dl \over 2 \pi} = -{c \over 6}\left[{2\over  |p|} \left( {1 \over 2}+{1 \over e^{\beta |p|}-1} \right) \right]+ {c \over 3} \log{\beta \over 2 \pi \epsilon} \delta(p) \, . 
\label{BE}
\ee
Remarkably the term on the rhs inside the square bracket reproduces the equal time Green's function of a free massless scalar at thermal equilibrium. Indeed, the second term in the parenthesis gives the Bose-Einstein statistics at inverse temperature $\beta$. We consider this an indication that the entanglement entropy is sensitive to the mode distribution of energy. The  dispersion relation $\epsilon(p)\!=\!|p|$ in \eqref{BE} could be directly related to the velocity $v^2\!=\!1$ with which the entanglement appears to propagate in a 2-dimensional CFT. 

Up to very suppressed effects, the action that brings the system out of equilibrium in the Vaidya model takes place for $t\!\in\! [-2a,2a]$. If entanglement propagates towards large scales with $v^2\!=\!1$, all entangled excitations at time $t$ will be separated at least a distance $2t\!-\!4a$.
This bound has also a holographic meaning: we have shown in \eqref{linearee} that geodesics associated to smaller intervals only perceive the black hole already formed. For $t\!>\!t_0$, with $t_0$ in \eqref{t0}, some of these geodesics are able to reach the black hole horizon and induce an extensive behavior for the cutoff independent piece of the entanglement entropy. This suggests that after $t_0$ regions of size $l\!<\!2t\!-\!4a$ have locally relaxed to a stationary state that can be identified as thermal equilibrium. Of course, in order to better sustain this statement it will be necessary the evaluation of other observables.

Larger intervals will contain entangled excitations. The holographic dictionary for the Vaidya model relates such intervals with geodesics that probe the black hole formation region, and for $l$ sufficiently large the empty AdS geometry previous to the collapse. Interestingly these geodesics present features absent in their BTZ counterpart. They cross behind the event horizon and moreover also behind the apparent horizon. The entanglement pattern among the sea of excitations and its causal propagation, determines that observables involving a length scale larger than $2t\!-\!4a$ at time $t$ can have an expectation value far from thermal. Hence equilibration cannot be achieved at the global level in a system of infinite size \cite{Calabrese:2005in,Cramer:2008zz}. We relate this property with the necessity to reach behind the apparent horizon in the holographic description of the field theory evolution. This behavior seems very generic for evolutions which admit a dual representation in terms of a process of gravitational collapse. 
Another observable able to feel that equilibration does not reach to regions of size $l\!\gtrsim\!2t$
are two-point functions, whose evolution after a quantum quench was studied in \cite{Calabrese:2006rx,Calabrese:2007rg}. It would be very interesting to explore the behavior of two-point functions in dynamical holographic models. \footnote{Based on semiclassical arguments, it has been proposed that the spacelike two-point function of operators with very high conformal dimension $\Delta$ is given holographically by $e^{-\Delta L}$, with $L$ the length of the geodesic that joints the two insertion points \cite{Banks:1998dd,Balasubramanian:1999zv}. For these reduced set of two-point functions, our analysis of geodesics implies consistency with the results of \cite{Calabrese:2006rx,Calabrese:2007rg}.}

Although we do not know the initial mode distribution created by the perturbation, the fact that the Vaidya metric only differs from AdS deep in the interior together with the absence of other characteristic scales in the problem, suggests that the typical momentum of the excitations sourced by the perturbation is set by the energy density $\epsilon(t)$ \eqref{energy}. This would imply that the energy is concentrated on modes of momentum up to $T\!=\!1/\beta$, with larger momenta being suppressed already at the moment of production. In order to describe entangled excitations as separated a certain distance,  it is necessary for the separation to be bigger than their wavelength. Therefore we can expect that this does not clearly happen before $t_0$, especially for short perturbations. On the gravity side, geodesics for $t\!<\!t_0$ enter the black hole formation region without necessarily crossing the apparent horizon. As $t$ decreases the apparent horizon influences ever less the shape of the geodesics. Moreover there is a time before which the geodesics never reach the apparent horizon, while the entanglement entropy already deviates appreciably from the vacuum result. For perturbations with a short time extent this happens well after the perturbation has ended. Hence the apparent horizon seems to lose relevance in the early stages of the evolution, as the picture of separated entangled excitations becomes increasingly inaccurate.

Besides $t_0$ an important time in the evolution of our holographic plasma is $\bar t$, after which the entanglement entropy enters a regime of steady evolution as described in \eqref{eeth}. 
We have argued based on \eqref{BE} that the entanglement entropy is sensitive to the mode distribution of energy. Assuming this, a possible explanation for the regime \eqref{eeth} could be that the mode occupation numbers have reached their final distribution. The only dynamics remaining on the system would be associated to the propagation of entanglement over ever larger scales, a process whose evolution is constrained by causality.
For perturbations with a short time extent, ${\bar t}$ is independent of $a$ and of the same order but larger than $t_0$, the time at which local thermalization can be first identified. Both facts suggest that \eqref{tth} is a characteristic time for the evolution of the holographic plasma, not directly related to the propagation of entanglement. This adds support to the interpretation of $\bar t$ in \eqref{tth} as a thermalization time for occupation numbers and would imply a very fast thermalization at strong coupling, on the line of related results for holographic plasmas \cite{Amado:2008ji,Chesler:2008hg,Beuf:2009cx,Chesler:2009cy}. At any rate, a firmer statement about the evolution of occupation numbers would require analyzing better adapted observables.

Finally we would like to show that a simple model similar to that proposed in \cite{Calabrese:2005in} to reproduce the evolution of entanglement entropy after a quantum quench for $l,t\!\gg\!0$ \eqref{eeevol}, can be generalized to account for the Vaidya evolution after ${\bar t}$. 
This model assumes that left and right moving entangled excitations propagate classically and without scattering, with a velocity $v^2\!=\!1$ in the case of a 2-dimensional CFT. Although this hypothesis could find support from a quasiparticle picture at weak coupling, such justification would not apply to strong coupling. However the evolution \eqref{eeevol} and therefore the success of the model is independent of the strength of the coupling. This suggests to drop any quasiparticle interpretation and consider the model as an effective way to describe the steady propagation of entanglement. Interestingly, the propagation of entanglement with velocity $v^2\!=\!1$ is supported not only by the study of dynamical setups, but by the result of Fourier transforming the entanglement entropy at thermal equilibrium \eqref{BE}. 

We generalize the model in \cite{Calabrese:2005in} by allowing for the presence of entanglement among excitations originating from points separated a distance $y$. For the sake of simplicity we will focus on a very brief perturbation, $a\!\simeq\!0$. We introduce a function $\rho(y)$ that describes the density of entangled excitations separated a distance $y$ at the moment of their production.  Given the translational invariance of the perturbations in our model, this function is independent of their location. In order to implement the steady regime \eqref{eeth}, we assume that at time $t$ the entanglement among degrees of freedom at a distance $y$ is described by
\be
\rho(y,t)= \left\{ \begin{array}{cl}
                           \rho(y-2t) & \,\,\,\,\, y\geq 2t  \, ,\\ 
                           0 & \,\,\,\,\, y<2t \, . \end{array} \right.
\label{rot}
\ee                           
Correlated excitations at $x$ and $x\!+\!y$ will contribute to the entanglement entropy of an interval when only one of these points lies inside the chosen interval
\be
S(l,t)=\int_{-\infty}^\infty dx \int_0^\infty dy \,(h(x)-h(x+y))^2 \rho(y,t) \, ,
\ee
with $h(x)\!=\!1$ if $x\!\in\![0,l]$ and zero otherwise. Substituting \eqref{rot}, we obtain an extensive result for intervals of size $l\!<\!2t$
\be
S(l,t)=2 l \int_0^\infty \rho(y) dy \, .
\ee
Coincidence with the cutoff independent term of \eqref{extee} determines $\int_0^\infty \rho dy=\pi c/ 6 \beta$. For $l\!>\!2t$ we then have
\be
S(l,t)= {2 \pi c t \over 3 \beta} 
+2(l-2t)\int_{l-2t}^\infty \rho(y) dy+\int_0^{l-2t}2 y \rho(y) dy \, ,
\label{eemodel}
\ee
in agreement with \eqref{eeth}. Moreover from the above relation we derive  
\be
\rho(y)=-{1 \over 2} \partial_l^2 S(l,t) \big|_{l=2t+y} \, .
\label{den}
\ee
The return of $S(l,t)$ to a logarithmic growth for sufficiently large intervals in the Vaidya model implies indeed the quantum entanglement between excitations produced with arbitrary large separations. Substituting \eqref{eeth} and \eqref{flarge} we obtain
\be
\rho(y)=-{1 \over 2} \partial_y^2 f(y)  \to {c \over 6 (y+x_0)^2} \, .
\ee
A preliminary analysis indicates that $x_0$ has only a mild dependence on the parameter $a$ and thus is approximately proportional to $\beta$. For the evolution after a quantum quench \eqref{eeevol}, relation \eqref{den} leads to $\rho(y)\!=\!{c \pi \over 6 \beta} \delta (y)$. Therefore we recover the initial localization of entanglement on very small scales and the picture in Fig.\ref{fig:prop}. 

In this paper we have explored the application of the AdS/CFT correspondence to the study of far from equilibrium dynamics in strongly coupled field theories, a very topical subject due to the heavy ion experiments at RHIC and LHC. We have focussed on the evaluation of the entanglement entropy, an observable with a simple holographic representation \cite{Ryu:2006bv,Ryu:2006ef}  generalizable to time-dependent situations \cite{Hubeny:2007xt}. A natural extension of our work would be to study the evolution of entanglement entropy for field theories in dimension higher than two, where few results are available. It would be also very interesting to evaluate the entanglement entropy in an equilibration process dual to the gravitational collapse of a shell with a more natural composition than null dust, as for example it was considered in \cite{Bhattacharyya:2009uu}. We plan to address some of the issues and open questions discussed in this section in future work.

%
\begin{acknowledgments}
We thank J. Barb\'on, C. G\'omez, R. Emparan, M. Garc{\'\i}a-P\'erez, D. Grumiller, V. Hubeny, K. Landsteiner, J. McGreevy, S. Minwalla, C. Pena, S. Ross, G. Sierra and S. Theisen for useful discussions and especially K. Landsteiner and G. Sierra for helpful comments on the draft. This work has been partially supported by grants FPA-2009-07908, HEPHACOS S2009/ESP-1473 and CPAN (CSD2007-00042). J. Abajo-Arrastia is supported by a FPU fellowship AP2007-01795. J. Apar{\'\i}cio is supported by the Portuguese Funda\c c\~ao para a Ci\^encia e Tecnologia, grant SFRH/BD/45988/2008.
\end{acknowledgments}

\bibliographystyle{JHEP}
\bibliography{therm}
\end{document}